\documentclass[%
 reprint,
 amsmath,amssymb,
 aps,
]{revtex4-2}

\usepackage{graphicx}
\usepackage{dcolumn}
\usepackage{bm}

\usepackage{color}     
\usepackage{amsmath}

\begin{document}

\preprint{APS/123-QED}

\title{Stellar $\beta$-decay rate of $s$-process branching-point $^{204}$Tl: forbidden transitions}

\author{Yang Xiao}%
\affiliation{School of Physical Science and Technology, Southwest University, Chongqing 400715, China}%

\author{Bin-Lei Wang}%
\affiliation{School of Physical Science and Technology, Southwest University, Chongqing 400715, China}%

\author{Long-Jun Wang}
\email{longjun@swu.edu.cn}
\affiliation{School of Physical Science and Technology, Southwest University, Chongqing 400715, China} 

\date{\today}

\begin{abstract}
  We propose a theoretical method to calculate the stellar $\beta$-decay rates of nuclei in stellar environments with high temperature and density, based on the projected shell model, where contributions from both allowed and first-forbidden transitions are taken into account. As the first example, the stellar $\beta$-decay rate of one of the last $s$-process branching-point nuclei, $^{204}$Tl, is calculated and studied, where all related transitions are first-forbidden transitions. For the terrestrial case, the ground-state to ground-state transition is unique first-forbidden transition, which is described reasonably by our calculations. At the typical $s$-process temperature ($T\approx 0.3$ GK), non-unique first-forbidden transitions from thermally populated excited states of the parent nucleus are involved, the effective rate from our calculations is much lower than the one from the widely used data tables by Takahashi and Yokoi. Effect of the quenching factors for nuclear matrix elements in first-forbidden transitions on the stellar $\beta$-decay rates is discussed as well.  
\end{abstract}

\maketitle


\section{\label{sec:intro}Introduction}

Nuclear $\beta$ decay plays crucial and indispensable roles in many aspects in nuclear astrophysics \cite{Fuller1980, Fuller1982_1, Fuller1982_2, fuller1985, langanke_RMP, langanke_2021_Rep_Pro_Phys}. The stellar $\beta$-decay rates in stellar environments with high temperature and high density, are important inputs for understanding many astrophysical problems regarding the evolution of stars and the origin of heavy elements, such as the core-collapse supernova of massive stars, the slow ($s$-) and rapid ($r$-) neutron-capture processes \cite{s_process_RMP_2011, r_process_RMP_2021}, the rapid proton-capture ($rp$-) process \cite{rp_process_Schatz_1998}, the cooling of neutron stars \cite{schatz2014nature} etc. In these problems, a large number of exotic nuclei with very short half-lives and/or transitions from excited states due to the thermal population are included, which are challenging for current experimental measurements. Therefore, theoretical calculations from different nuclear many-body methods are relied on heavily, and reliable nuclear inputs become increasingly important. 

The $s$ process contributes approximately half of the solar elemental abundances between Fe and Bi \cite{s_process_RMP_2011}. The $s$ process is expected to take place during core He-burning and shell C-burning of massive stars (weak $s$ process) as well as in H-burning and He-burning layers of the Asymptotic Giant Branch (AGB) phase of low-mass stars (main $s$ process), with the typical temperature (neutron and electron density) reaching $\approx 0.3$ GK ($\approx 10^{10}$ cm$^{-3}$ and $\approx 10^{26}$ cm$^{-3}$ respectively) \cite{s_process_RMP_2011, TY_table_1987, Lugaro_Review_2023_s_beyond}. An important and interesting problem during the $s$ process is the branching point. It refers to unstable isotope for which the neutron-capture reaction rate can compete with corresponding $\beta$-decay rate \cite{s_process_RMP_2011, diehl2018astrophysics}. The branching-point nuclei play decisive roles in determining the reaction flow, the path, the final abundances etc. due to the competition between the two branches, further neutron capture and $\beta^-$ decay, where the effective $\beta$-decay rates in $s$-process environments are indispensable inputs.

\begin{figure}[htbp]
\begin{center}
  \includegraphics[width=0.48\textwidth]{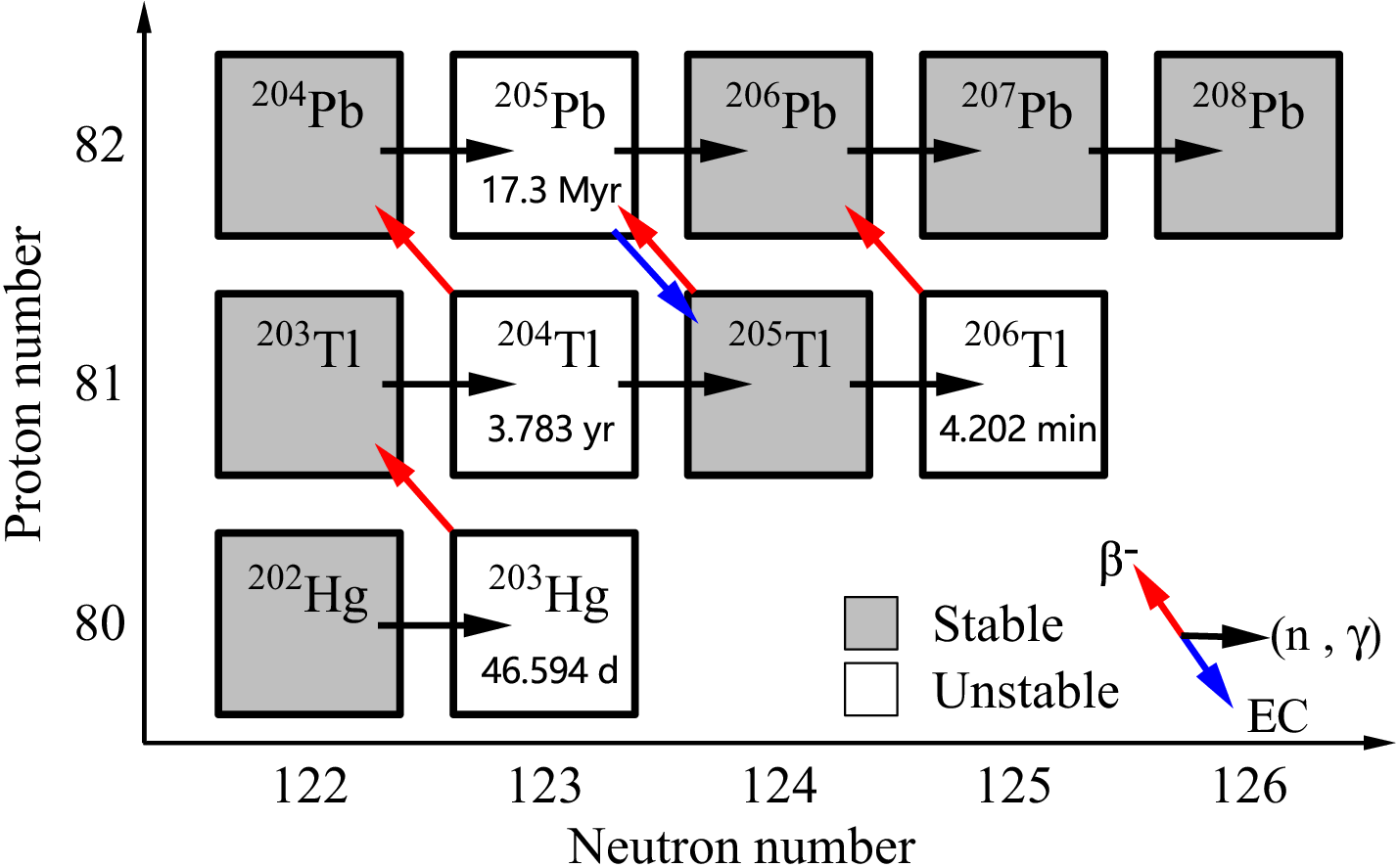}
  \caption{\label{fig:fig1} (color online) An excerpt of the chart of nuclides near the termination of the $s$ process, where the branching-point $^{204}$Tl is illustrated. Neutron capture, $\beta$ decay and electron capture (EC) are labeled by arrows. }
\end{center}
\end{figure}

An example of the $s$-process branching points, $^{204}$Tl, is shown in Fig. \ref{fig:fig1}. On one hand, $^{204}$Tl is one of the last two branching points near the termination of the $s$ process (the other is $^{205}$Tl for which bound-state $\beta^-$ decay is open when temperature $T \gtrsim 0.16$ GK) \cite{TY_table_1987, Doming0_PRC_2007, Ratzel_PRC_2004, Broda_PRC_2011_high_spin, Utsunomiya_PRC_2019, Casanovas_2020_JP_Con, Casanovas_2018_EPJ_Web, Casanovas_2022_EPJ_Web}. The relevant nuclei, $^{204}$Pb and $^{205}$Pb, are $s$-only nuclei which can be produced only by the $s$ process, so that they are potential chronometers of the last $s$-process events and they become probes of the corresponding stellar conditions. It is seen from Fig. \ref{fig:fig1} that the $\beta^-$ decay of $^{204}$Tl is crucial to determine the final abundance of $^{204}$Pb. On the other hand, the stellar $\beta^-$-decay rate of $^{204}$Tl is predicted to depend sensitively on temperature, leading to $^{204}$Tl as an $s$-process thermometer. Besides, nuclear allowed transitions usually dominate the stellar $\beta^-$ decays of branching points, while in many cases such as for $^{151}$Sm, $^{170}$Tm, $^{181}$Hf, $^{185}$W, $^{204}$Tl etc. the stellar $\beta^-$ decay rates are determined by first-forbidden transitions, where $^{204}$Tl serves as one of the special examples as all the related transitions are first-forbidden. 

Generally the stellar decay rates from the data tables by Takahashi and Yokoi (referred to as TY87 hereafter) \cite{TY_table_1987} are adopted for studies of $s$ process (branching points). The TY87 data are outdated as the corresponding comparative half-life, $ft$, of transitions from excited states are determined neither from experimental measurements nor from sophisticated nuclear many-body calculations, but are actually estimated empirically from systematics of transitions of neighboring nuclei with available data (usually from ground states). The TY87 data regarding transitions from excited states have been challenged by modern charge-exchange reactions (for $^{59}$Fe \cite{BGao_2021_PRL_59Fe}) or shell-model calculations (for $^{134}$Cs \cite{Kuo_Ang_Li_2021_ApJL}) for allowed transitions recently. Over the past decade, the projected shell model (PSM) \cite{PSM_review, Sun_1996_Phys_Rep} is extended to have large configuration space considering high-order quasiparticle (qp) configurations with the help of the Pfaffian and other algorithms \cite{Mizusaki_2013_PLB, LJWang_2014_PRC_Rapid, LJWang_2016_PRC, ZRChen_2022_PRC, BLWang_2022_PRC}. The PSM is then applied to studies on nuclear high-spin physics \cite{LJWang_2016_PRC, LJWang_2019_JPG, LJWang_PLB_2020_chaos, Petrache_PRC_2023}. Besides, the PSM is further developed for describing allowed Gamow-Teller (GT) transition and stellar weak-interaction rates \cite{Z_C_Gao_2006_GT, LJWang_2018_PRC_GT, LJWang_PLB_2020_ec, LJWang_2021_PRL, LJWang_2021_PRC_93Nb, zrchen2023symm, ZRChen_PLB2023} and nuclear $\beta$ spectrum \cite{FGao_PRC2023} recently. A new development of the PSM for description of first-forbidden transition of nuclear $\beta$ decay is accomplished very recently \cite{BLWang_1stF_2024}. It is then straightforward to present a PSM method for calculating the stellar weak-interaction rates taking into account both allowed GT and first-forbidden transitions, and taking the branching-point $^{204}$Tl as an example for applications here. 

The paper is organized as follows. In Sec. \ref{sec:framework} we introduce the basic framework for stellar $\beta$-decay rates taking into account both allowed and first-forbidden transitions. The calculations and predictions for the stellar $\beta^-$ decay rate of $^{204}$Tl are discussed in details in Sec. \ref{sec:result} and we finally summarize our work in Sec. \ref{sec:sum}.

\section{\label{sec:framework}Methodology for stellar $\beta$-decay rates}

Within the common assumption that parent nuclei are in a thermal equilibrium with occupation probability for excited states following the Boltzmann distribution, the stellar $\beta^-$ decay rates read as \cite{Fuller1980, Fuller1982_1, Fuller1982_2, fuller1985},
\begin{eqnarray} \label{eq.total_lambda}
  \lambda^{\beta^{-}} = \sum_{if} \frac{(2J_{i}+1)e^{-E_{i}/(k_{B}T)}}{G(Z,A,T)} \lambda^{\beta^-}_{if},
\end{eqnarray}
where the summations run over initial ($i$) and final ($f$) states of parent and daughter nuclei, respectively (with spin-parity assignments $J_i^{\pi_i}, J_f^{\pi_f}$ and excitation energies $E_i, E_f$), $k_{B}$ labels the Boltzmann constant and $T$ is the environment temperature. $G(Z, A, T) = \sum_i (2J_i + 1)\text{exp}(-E_{i}/(k_{B}T))$ is the partition function. The factor in Eq. (\ref{eq.total_lambda}) reflects the thermal population probability of the initial (low-lying) state. The individual rate,
\begin{align} \label{eq.lambda_if}
  \lambda^{\beta^-}_{if} 
  =& \frac{\ln 2}{K} \int_{1}^{Q_{if}} C(W) F_0(Z+1, W) pW (Q_{if} - W)^{2} \nonumber \\
     & \qquad \quad \times (1 - S_e(W)) dW,
\end{align}
where the constant $K$ can be determined from superallowed Fermi transitions and $K=6144 \pm 2$ s \cite{Hardy_2009_PRC} is adopted in this work. $W$ and $p = \sqrt{W^2 - 1}$ label the total energy (rest mass and kinetic energy) and the momentum of the electron in units of $m_e c^2$ and $m_e c$ respectively. The available total energy for leptons in the one-to-one $\beta^-$ transition is given by,
\begin{eqnarray} \label{eq.Qif}
  Q_{if} = \frac{1}{m_e c^2 }(M_p - M_d + E_i -E_f ), 
\end{eqnarray}
where $M_p (M_d)$ indicates the nuclear mass of parent (daughter) nucleus. In Eq. (\ref{eq.lambda_if}), $S_e$ is the electron distribution function following the Fermi-Dirac distribution as,
\begin{eqnarray} \label{eq.Se}
  S_e (W) = \frac{1}{\text{exp}[(W - \mu_e) / k_B T] + 1} ,
\end{eqnarray}	
in which the chemical potential, $\mu_{e}$, is determined from the relation,
\begin{eqnarray} \label{eq.mue}
  \rho Y_e = \frac{1}{\pi^2 N_A} \left( \frac{m_e c}{\hbar} \right)^3 \int_{0}^{\infty} (S_e - S_p) p^2 dp,
\end{eqnarray}
here $N_A$ represents Avogadro's number, and $\rho Y_e$ labels the electron density. Note that one can get the positron distribution $S_p$ by the replacement $\mu_{p} = -\mu_{e}$. In Eq. (\ref{eq.lambda_if}) $F_0(Z+1, W)$ is the Fermi function that accounts for the Coulomb distortion of the electron wave function near the nucleus \cite{Fuller1980, Fermi_func_1983}, with $Z$ being the proton number of the decaying (parent) nucleus. 

The $C(W)$ term in Eq. (\ref{eq.lambda_if}) is the shape factor for the decay. In the case of allowed transitions with selection rule $|\Delta J| = |J_i - J_f| = 0,1$, and $\Delta \pi=\pi_i \pi_f = +1$, $C(W)$ does not depend on the electron energy \cite{Zhi_FF_PRC_2013}, and can be written as,
\begin{eqnarray} \label{eq.BGT_if}
  C(W) = B(\text{GT}^-)_{if} =
  \left(\frac{g_A}{g_V}\right)^2_{\text{eff}} 
  \frac{\big\langle \Psi _{J_f}^{n_{f}} \big\| \sum_{k} \hat{\bm\sigma}^k \hat\tau_{-}^k \big\| \Psi_{J_i}^{n_i} \big\rangle^2 }{2J_{i}+1} , \nonumber \\
\end{eqnarray}
where only GT contribution is considered as usual since the contribution coming from the Fermi operator is usually negligible \cite{Cole_2012_PRC, Sarriguren_2013_PRC, Martinez_Pinedo_2014_PRC}. Here $\hat{\bm\sigma}$ ($\hat\tau_{-}$) is the Pauli spin operator (isospin lowering operator), the summation runs over all nucleons, and $\Psi_{J}^{n}$ represents the $n$-th eigen nuclear many-body wave function for angular momentum $J$.  $(g_{A}/g_{V})_{\text{eff}}$ is the effective ratio of axial and vector coupling constants with corresponding quenching for the GT matrix element (transition operator and/or nuclear wave function) \cite{A.brown1985, martinez1996, Gysbers_2019_Nat_Phys, Javier2011PRL, LJWang_current_2018_Rapid}, 
\begin{eqnarray} \label{eq.quench_GT}
  \left( \frac{g_{A}}{g_{V}}\right)_{\text{eff}} = f_q(\text{GT}) \left(\frac{g_{A}}{g_{V}} \right)_{\text{bare}} ,
\end{eqnarray}
where $(g_A/g_V)_{\text{bare}} = -1.27641(45)$ \cite{arkisch2019}, and $f_q(\text{GT})$ is the quenching factor for GT transition.

For first-forbidden transitions, $C(W)$ has explicit energy dependence, which is approximated as \cite{Weidenmuller_FF_RMP_1961, Zhi_FF_PRC_2013, Mougeot_PRC_2015}
\begin{eqnarray} \label{eq.CW}
  C(W) = k(1 + aW + b/W + cW^2) ,
\end{eqnarray}
for non-unique transitions with $|\Delta J| = 0,1$, $\Delta \pi = -1$, and unique transitions with $|\Delta J|=2$, $\Delta\pi=-1$. The coefficients $k, a, b$ and $c$ depend on the nine reduced nuclear matrix elements in Eq.(\ref{eq.all_ME}) for first-forbidden transition in the following way (see Refs. \cite{Zhi_FF_PRC_2013, BLWang_1stF_2024} for details), 
\begin{eqnarray} \label{eq.kabc}
  k  &=& [\zeta_0^2 + \frac{1}{9} w^2]^{(0)} + [\zeta_1^2 + \frac{1}{9}(x+u)^2 - \frac{4}{9}\mu_1\gamma_1 u(x+u)   \nonumber \\
     & & \quad + \frac{1}{18} Q_{if}^2 (2x+u)^2 - \frac{1}{18} \lambda_2 (2x-u)^2 ]^{(1)} \nonumber \\
     & & \quad + [\frac{1}{12} z^2 (Q_{if}^2-\lambda_2)]^{(2)} , \nonumber \\
  ka &=& [-\frac{4}{3} u Y -\frac{1}{9} Q_{if} (4x^2 +5u^2)]^{(1)} -[\frac{1}{6} z^2 Q_{if}]^{(2)} , \nonumber \\
  kb &=& \frac{2}{3} \mu_1 \gamma_1 \{ -[\zeta_0 w]^{(0)} + [\zeta_1 (x+u)]^{(1)} \} , \nonumber \\
  kc &=& \frac{1}{18} [8u^2 +(2x+u)^2 + \lambda_2 (2x-u)^2]^{(1)} \nonumber \\
     & & \quad +\frac{1}{12} [z^2 (1+\lambda_2)]^{(2)} ,
\end{eqnarray}
where the numbers in parentheses of superscripts denote the rank of the transition operators, $\xi=\alpha Z/(2R)$ with $\alpha$ and $R$ being the fine structure constant and the radius of the nucleus respectively. $\gamma_1=\sqrt{1-(\alpha Z)^2}$ and $\mu_1\approx 1$ are adopted as in Refs. \cite{Zhi_FF_PRC_2013, BLWang_1stF_2024}. $\lambda_2 = \frac{F_1(Z, W)}{F_0(Z, W)}$ with $F_1$ being the generalized Fermi function \cite{Suhonen_PRC_2017_general_Fermi_for_unique}. In Eq. (\ref{eq.kabc}) $\zeta_0, \zeta_1$ and $Y$ are defined as follows, 
\begin{align} \label{eq.VY}
  V &= \xi' \nu + \xi w',   \qquad \ \quad \zeta_0 = V + \frac{1}{3} w Q_{if}, \nonumber \\
  Y &= \xi'y - \xi (u'+x'), \quad  \zeta_1 = Y + \frac{1}{3} (u-x) Q_{if} .  
\end{align}

In Eqs. (\ref{eq.kabc}, \ref{eq.VY}) the involved nuclear matrix elements read as, 
\begin{subequations} \label{eq.all_ME}
\begin{eqnarray}
  w  &=& -g_A \sqrt{3} \frac{\left\langle \Psi^{n_f}_{J_f} \left\| \sum_k r_k [\bm C^k_1 \otimes \hat{\bm\sigma}^k]^0 \hat{\tau}^k_- \right\| \Psi^{n_i}_{J_i} \right\rangle}{\sqrt{2J_i+1}} , \\
  x  &=& - \frac{\left\langle \Psi^{n_f}_{J_f} \left \| \sum_k r_k \bm C^k_1 \hat{\tau}^k_- \right \| \Psi^{n_i}_{J_i} \right \rangle}{\sqrt{2J_i+1}} , \\
  u  &=& -g_A \sqrt{2} \frac{\left\langle \Psi^{n_f}_{J_f} \left \| \sum_k r_k [\bm C^k_1 \otimes \hat{\bm\sigma}^k]^1 \hat\tau^k_- \right \| \Psi^{n_i}_{J_i} \right \rangle}{\sqrt{2J_i+1}} , \\
  z  &=& 2g_A \frac{\left\langle \Psi^{n_f}_{J_f} \left \| \sum_k r_k [\bm C^k_1 \otimes \hat{\bm\sigma}^k]^2 \hat\tau^k_- \right \| \Psi^{n_i}_{J_i} \right \rangle}{\sqrt{2J_i+1}} , \\ 
  w' &=& -g_A \sqrt{3} \frac{\left\langle \Psi^{n_f}_{J_f} \left \| \sum_k \frac{2}{3}r_k I(r_k) [\bm C^k_1 \otimes \hat{\bm\sigma}^k]^0 \hat\tau^k_- \right \| \Psi^{n_i}_{J_i} \right \rangle}{\sqrt{2J_i+1}} , \nonumber \\ \\
  x' &=& - \frac{\left\langle \Psi^{n_f}_{J_f} \left \| \sum_k \frac{2}{3}r_k I(r_k)  \bm C^k_1 \hat\tau^k_- \right \| \Psi^{n_i}_{J_i} \right \rangle}{\sqrt{2J_i+1}} , \\
  u' &=& -g_A \sqrt{2} \frac{\left\langle \Psi^{n_f}_{J_f} \left \| \sum_k  \frac{2}{3}r_k I(r_k) [\bm C^k_1 \otimes \hat{\bm\sigma}^k]^1 \hat\tau^k_- \right \| \Psi^{n_i}_{J_i} \right \rangle}{\sqrt{2J_i+1}} , \nonumber \\ \\
  \xi'\nu &=& \frac{g_A\sqrt{3}}{M_0} \frac{\left\langle \Psi^{n_f}_{J_f} \left \| \sum_k  [\hat{\bm\sigma}_k \otimes \bm\nabla^k]^0 \hat\tau^k_- \right \| \Psi^{n_i}_{J_i} \right \rangle}{\sqrt{2J_i+1}} , \\
  \xi' y  &=& - \frac{1}{M_0} \frac{\left\langle \Psi^{n_f}_{J_f} \big\| \sum_k  \bm\nabla^k \hat\tau^k_- \big\| \Psi^{n_i}_{J_i} \right \rangle}{\sqrt{2J_i+1}}  .
\end{eqnarray} 
\end{subequations}
where the definitions of $\bm C_{lm}$, the radial function $I(r)$ etc. can be seen from Refs. \cite{Zhi_FF_PRC_2013, BLWang_1stF_2024}. 

As in the allowed GT transition of $\beta$ decay \cite{A.brown1985, martinez1996, Gysbers_2019_Nat_Phys} and in the double $\beta$ decay \cite{Javier2011PRL, LJWang_current_2018_Rapid}, quenching factors for nuclear matrix elements (transition operators and/or nuclear wave functions) of first-forbidden transitions in Eq. (\ref{eq.all_ME}) should probably be introduced as well \cite{Zhi_FF_PRC_2013}. In this work the following quenching factors are adopted,
\begin{align} \label{eq.quench}
  f_q(\xi'\nu) &= 1.266,          \hspace{3.95em} f_q(w) = f_q(w') = 0.66, \nonumber \\
  f_q(x)       &= f_q(x') = 0.51, \quad f_q(u) = f_q(u') = 0.38, \nonumber \\
  f_q(z)       &= 0.867 . 
\end{align}
which are similar to the ones adopted in shell-model studies in Ref. \cite{Zhi_FF_PRC_2013} except that $f_q(z)$ is reevaluated from 0.42 to 0.867 by fitting the data for transition between ground states as shown in Fig. \ref{fig:fig3}.

\section{\label{sec:result}Analysis for stellar $\beta^-$ decay rates of $^{204}$Tl}

For both the allowed transitions as in Eq. (\ref{eq.BGT_if}) and the first-forbidden transitions as in Eq. (\ref{eq.all_ME}), theoretical description depends on reliable calculations of corresponding nuclear matrix elements from elaborate nuclear models where nuclear many-body wave functions in the laboratory frame, $\Psi_{J}^{n}$, should be prepared first. In this work, $\Psi_{J}^{n}$ is described by the PSM \cite{LJWang_2014_PRC_Rapid, LJWang_2016_PRC} based on the angular-momentum-projection techniques as,
\begin{eqnarray} \label{eq.wave_function}
  | \Psi^{n}_{JM} \rangle = \sum_{K\kappa} f_{K\kappa}^{Jn} \hat{P}_{MK}^{J} | \Phi_{\kappa} \rangle ,
\end{eqnarray}
where $\Phi_{\kappa}$ labels the qp vacuum and different qp excitation configurations \cite{LJWang_2014_PRC_Rapid, LJWang_2018_PRC_GT} and the angular-momentum-projection operator reads as,
\begin{eqnarray} \label{AMP_operator}
    \hat{P}^{J}_{MK} = \frac{2J + 1}{8\pi^2} \int d\Omega D^{J\ast}_{MK} (\Omega) \hat{R} (\Omega) ,
\end{eqnarray}
where $D^J_{MK}$ ($\hat R$) is the Wigner $D$ function (rotation operator) with respect to the Euler angle $\Omega$ \cite{ZRChen_2022_PRC, BLWang_2022_PRC} with $M$ ($K$) being the spin projection in the laboratory (intrinsic) frame. $f$ in Eq. (\ref{eq.wave_function}) labels the expansion coefficients that can be obtained by solving the corresponding eigen equation. The projection operator transforms the description of nuclei from the intrinsic to the laboratory frame so that different physical quantities of interests can be calculated and compared with measurements straightforwardly. The detailed expressions, algorithms and numerical implementation for related nuclear matrix elements for allowed GT transitions as in Eq. (\ref{eq.BGT_if}) are accomplished by the PSM with Eq. (\ref{eq.wave_function}) in recent years \cite{LJWang_2018_PRC_GT}, and those for first-forbidden transitions as in Eq. (\ref{eq.all_ME}) are accomplished by us very recently \cite{BLWang_1stF_2024}. Here for brevity we avoid repeating these details for which one can refer to Refs. \cite{LJWang_2018_PRC_GT, BLWang_1stF_2024}. In this work we actually implemented the calculations of nuclear matrix elements for both allowed GT and first-forbidden transitions into the framework in Sec. \ref{sec:framework} so that a practical PSM model for stellar $\beta$-decay rates taking into account both allowed and first-forbidden transitions is obtained numerically. The $s$-process branching-point $^{204}$Tl is taken as the first example for applications and discussions.

\begin{figure}[t]
\begin{center}
  \includegraphics[width=0.48\textwidth]{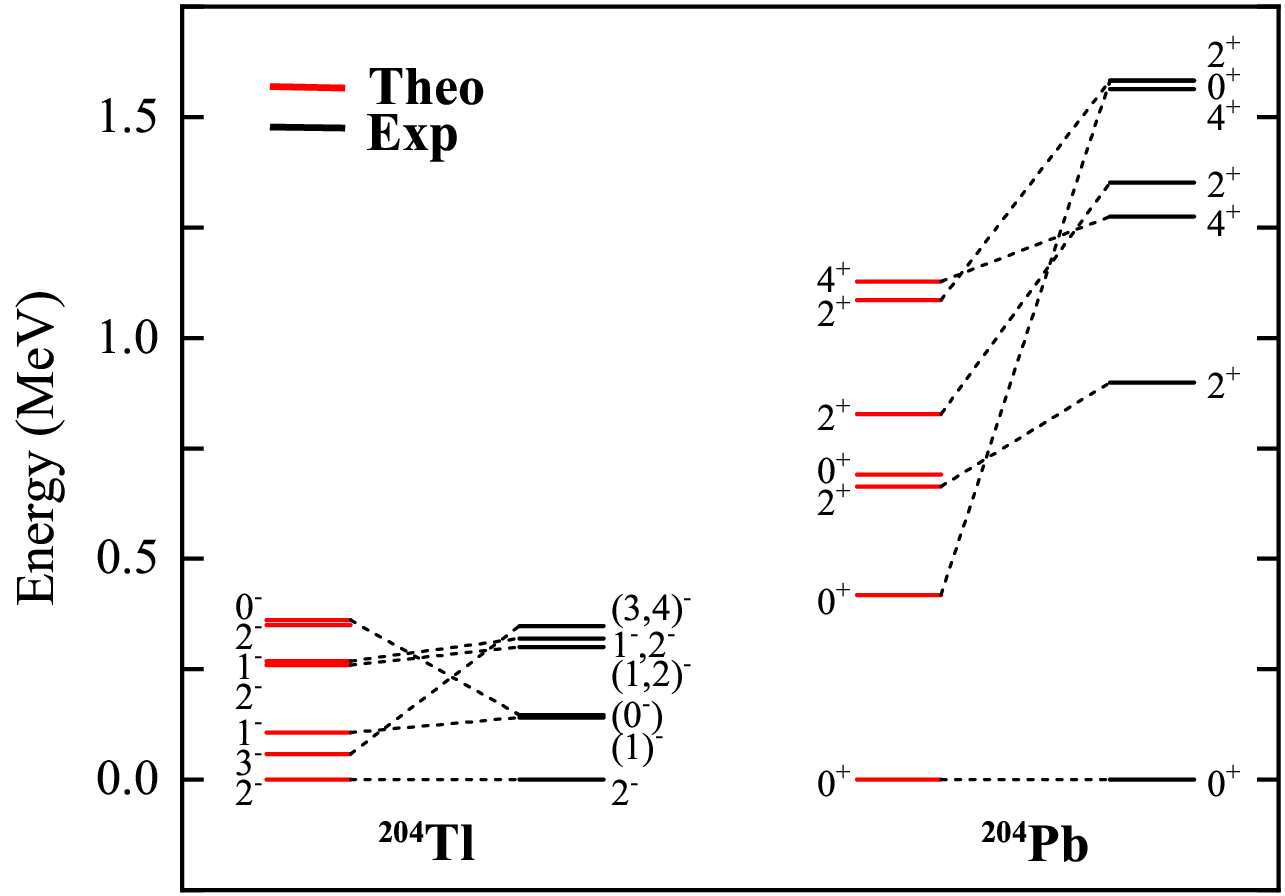}
  \caption{\label{fig:fig2} (color online) The calculated excitation energies for low-lying states of the parent and daughter nuclei, as compared with available data \cite{NNDC}. }
\end{center}
\end{figure}

\begin{figure}[t]
\begin{center}
  \includegraphics[width=0.48\textwidth]{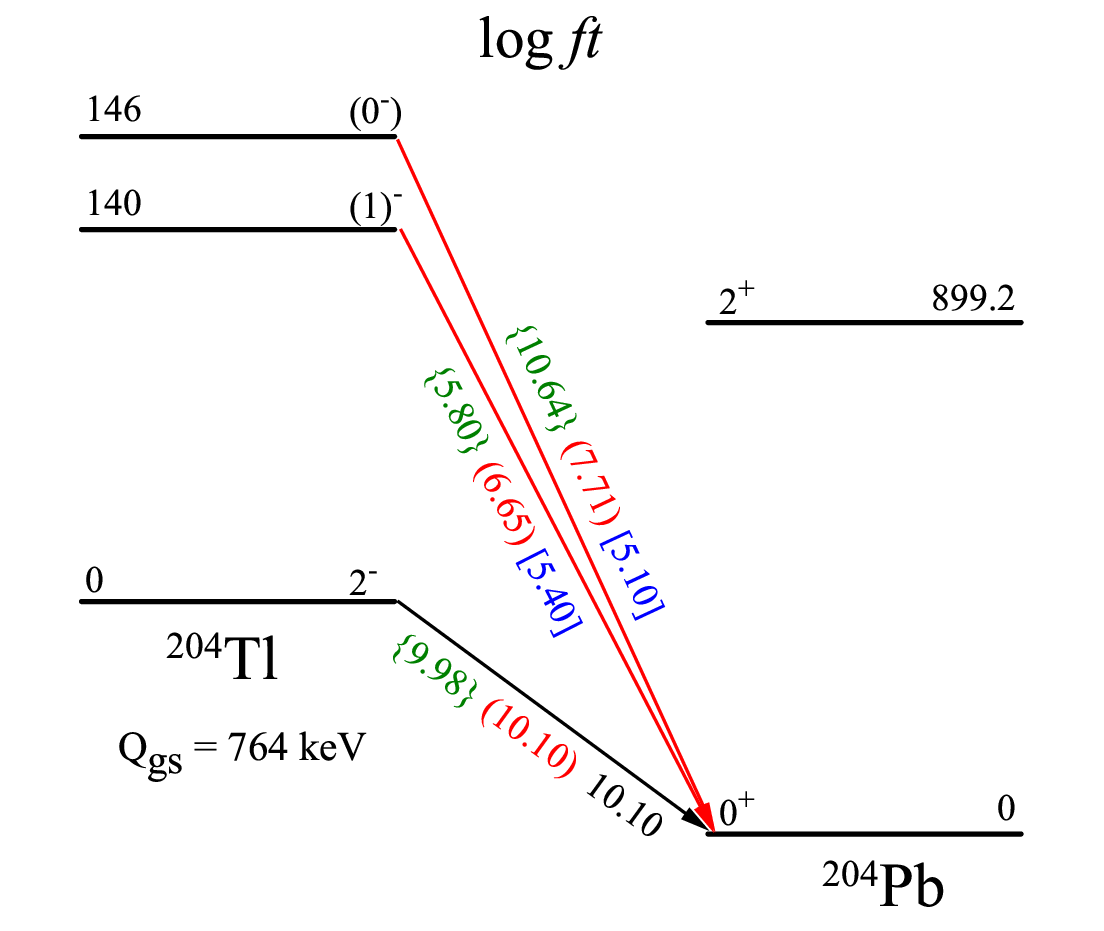}
  \caption{\label{fig:fig3} (color online) The schematic stellar $\beta$-decay scheme of $^{204}$Tl. Calculated log$ft$ values without (in green brace) and with (in red parenthesis) the quenching factors in Eq. (\ref{eq.quench}) are compared with the TY87 results (in blue bracket) \cite{TY_table_1987} for transitions from low-lying states, and the data \cite{NNDC} for transition between ground states (gs). Excitation energies are in keV. }
\end{center}
\end{figure}

Figure \ref{fig:fig2} illustrates the calculated energy levels for the parent $^{204}$Tl and daughter $^{204}$Pb nuclei as compared with the data \cite{NNDC}. It is seen that the data are described reasonably. In particular, the spin and parity of the ground state for the odd-odd nucleus $^{204}$Tl is well reproduced. Besides, the energy gap between the ground state and the excited states for $^{204}$Pb is described reasonably, and the behavior of dense levels for low-lying states of $^{204}$Tl is presented. The first and second excited states of $^{204}$Tl are preset tentatively to have spin-parity $1^-$ and $0^-$, which is supported by our PSM calculations. The experimental excitation energies for these two low-lying states are 140 keV and 146 keV respectively, which are difficult to be thermally populated effectively at the typical $s$-process temperature $T\approx 0.3$ GK. However, provided that the $\beta$-decay $Q$ value of $^{204}$Tl is as small as 764 keV, and because that the transitions from these two low-lying states are predicted in the TY87 data tables \cite{TY_table_1987} to be much stronger than the terrestrial transition between ground states by about five orders of magnitude, as seen from Fig. \ref{fig:fig3}, even tiny thermal population probability of them could increase rapidly the stellar $\beta$ decay rate of $^{204}$Tl with the temperature, so that leading to $^{204}$Tl as an $s$-process thermometer and affecting the potential chronometers of the last $s$-process events, $^{205}$Pb.

Due to the importance as discussed above, it is necessary to study the stellar $\beta$-decay rate of $^{204}$Tl within elaborate nuclear many-body method, since the unknown transition strengths from excited states are estimated empirically in the  TY87 data tables \cite{TY_table_1987}. Figure \ref{fig:fig3} shows the schematic $\beta$-decay scheme of $^{204}$Tl in the $s$-process environments. In terrestrial case, the $^{204}$Tl stays in the ground state and its half-life is 3.783 yr, which is dominated by the $\beta^-$-decay channel with the EC channel contributing less than 3$\%$, so that we focus on the $\beta^-$-decay channel in the following. Owing to the small $Q$ value, only ground-state to ground-state transition is relevant for the $\beta^-$ decay of $^{204}$Tl to $^{204}$Pb in terrestrial case. Such a transition is unique first-forbidden and is found experimentally to be much weak with log$ft=10.10$. It is seen from Fig. \ref{fig:fig3} that our PSM calculations present stronger transition strength than the data with log$ft = 9.98$ when no quenching factor is adopted. The data can be described reasonably by our PSM calculations when the quenching factors in Eq. (\ref{eq.quench}) are adopted.

\begin{figure}[t]
\begin{center}
  \includegraphics[width=0.49\textwidth]{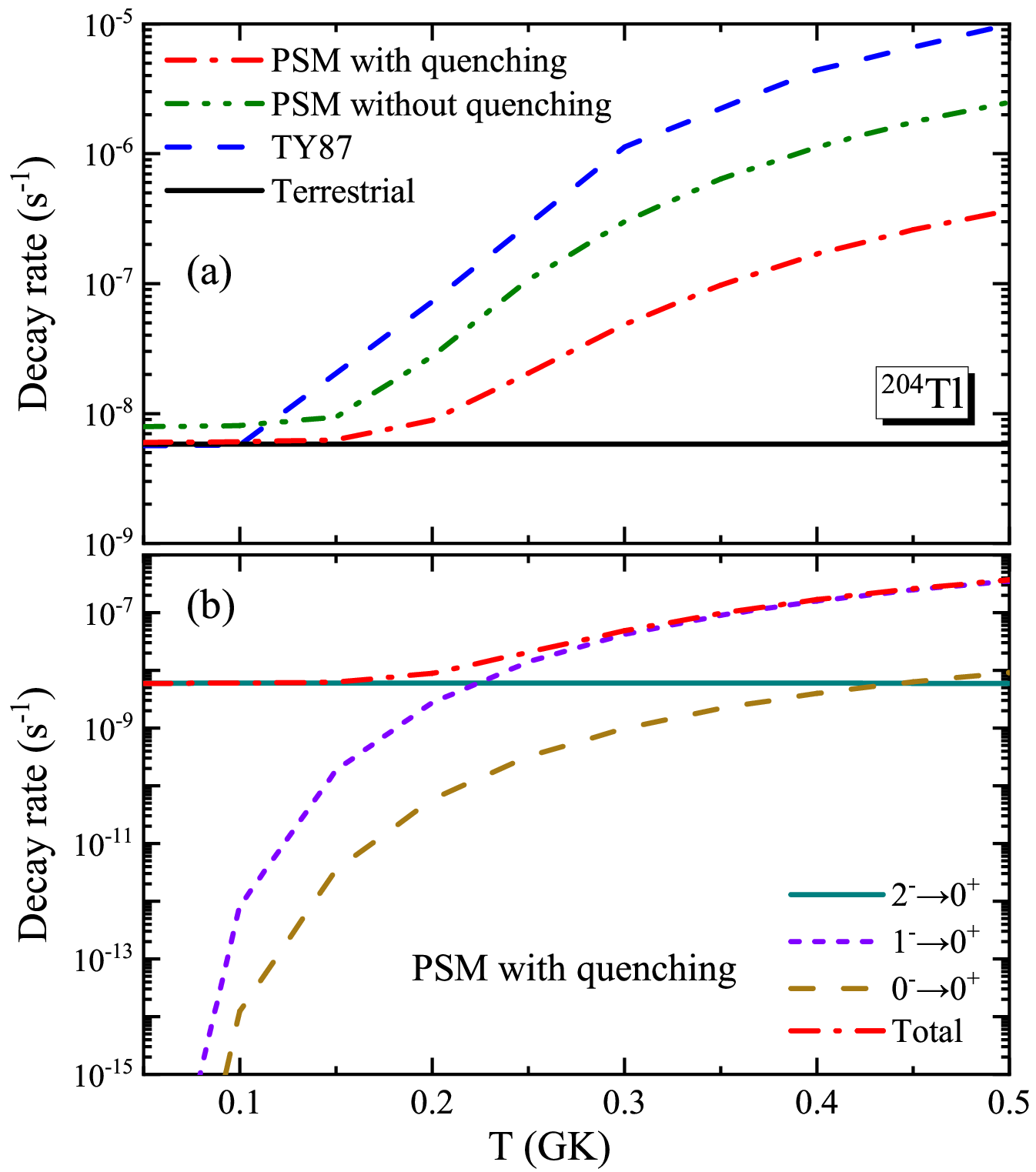}
  \caption{\label{fig:fig4} (color online) (a) The calculated stellar $\beta^-$-decay rate of $^{204}$Tl by the PSM with and without the quenching factors in Eq. (\ref{eq.quench}) as a function of the stellar temperature, compared with the TY87 rate \cite{TY_table_1987} and the constant terrestrial rate. (b) The partial decay rates of individual transitions by the PSM calculations with the quenching factors, compared to the total one. See the text for details. }
\end{center}
\end{figure}

With the increase of temperature during the $s$ process, the first and second excited states of $^{204}$Tl may be populated thermally. The transitions from these two low-lying states then become relevant, which are non-unique first-forbidden and the transition strengths are expected to be much strong with log$ft \approx 5$ by the TY87 data tables \cite{TY_table_1987} as seen from Fig. \ref{fig:fig3}. By comparison, the corresponding transition strengths predicted by the PSM calculations are reduced largely and turn out to be dependent sensitively on the quenching. On one hand, this indicates that the stellar $\beta^-$-decay rate of $^{204}$Tl in the $s$-process environments may be reduced largely than the one from the widely used TY87 data tables. On the other hand, the quenching effect of nuclear matrix elements for first-forbidden transitions is important in our understanding of the roles played by first-forbidden transitions in the stellar weak-interaction rates.

With the experimental energy levels and the calculated shape factors $C(W)$ for the three first-forbidden transitions in Fig. \ref{fig:fig3} by the PSM, the calculated stellar $\beta^-$-decay rate of $^{204}$Tl is compared with the TY87 rate and the terrestrial one in Fig. \ref{fig:fig4}(a), where the electron number density is adopted as $\rho Y_e = 1.66\times 10^2$ mol/cm$^{3}$ in the PSM calculations (this corresponds to $n_e = 10^{26}$ cm$^{-3}$ in Ref. \cite{TY_table_1987} and the rate of $^{204}$Tl shows almost no sensitive dependence on electron density in our calculations and in Ref. \cite{TY_table_1987}). It is seen that as the $0^- \rightarrow 0^+$ transition is evaluated to be stronger than the ground-state to ground-state one by five orders of magnitude (Fig. \ref{fig:fig3}), the TY87 rate increases very rapidly when the temperature $T \geqslant 0.1$ GK. Such a sensitive temperature dependence makes $^{204}$Tl as a potential $s$-process thermometer. By comparison, the calculated rate by the PSM without (with) the quenching factors in Eq. (\ref{eq.quench}) is higher than (equivalent to) the terrestrial data at $T \lesssim 0.1$ GK, owing to that the transition between ground states is calculated to be stronger than (equivalent to) the data as seen from Fig. \ref{fig:fig3}. Since the transitions from two related low-lying states are predicted by the PSM to be weaker than the TY87 results, at the $s$-process typical temperature $T \approx 0.3$ GK, the predicted rate by the PSM calculations is much reduced than the corresponding TY87 rate by about a factor of five (1.5 orders of magnitude) without (with) the quenching of nuclear matrix elements. The stellar $\beta^-$-decay rate of $^{204}$Tl is then probably much lower than what is expected before, and the effect of the quenching on the decay rate could be as large as about one order of magnitude. It is worth mentioning that with the experimental developments of neutron-capture cross sections, uncertainty in stellar $\beta$-decay rates need to be reduced with reliable nuclear many-body models, when first-forbidden transition play important roles, the corresponding quenching problem needs further investigation.

Finally, the partial decay rates from the three individual transitions in Fig. \ref{fig:fig3}, i.e., the $\lambda^{\beta^-}_{if}$ weighted by corresponding thermal population probability of initial states as shown in Eq. (\ref{eq.total_lambda}), by the PSM calculations with the quenching factors, are plotted in Fig. \ref{fig:fig4}(b) along with the total one. At low temperature, the total rate is dominated by the transition between ground states as the parent nucleus stays in its ground states. The partial decay rate for transition between ground states keeps nearly constant with the increase of the temperature $T$, indicating that only tiny population probability is shared by the ground state to low-lying state for $^{204}$Tl at $s$-process environments. In practice, at the peak temperature with $T \approx 0.5$ GK, only about $3\%$ population probability is shared, which is dominated by the $1^- \rightarrow 0^+$ transition (about $2.3\%$) due to the large degeneracy ($2J_i+1$) of the $1^-$ state compared with the one of the $0^-$ state. Besides, such transition is predicted to be stronger than the $0^- \rightarrow 0^+$ transition with log$ft=6.65$ as seen from Fig. \ref{fig:fig3}, so that the total decay rate is dominated by the $1^- \rightarrow 0^+$ transition at high temperature, as seen from Fig. \ref{fig:fig4}(b).

\begin{figure}[t]
\begin{center}
  \includegraphics[width=0.46\textwidth]{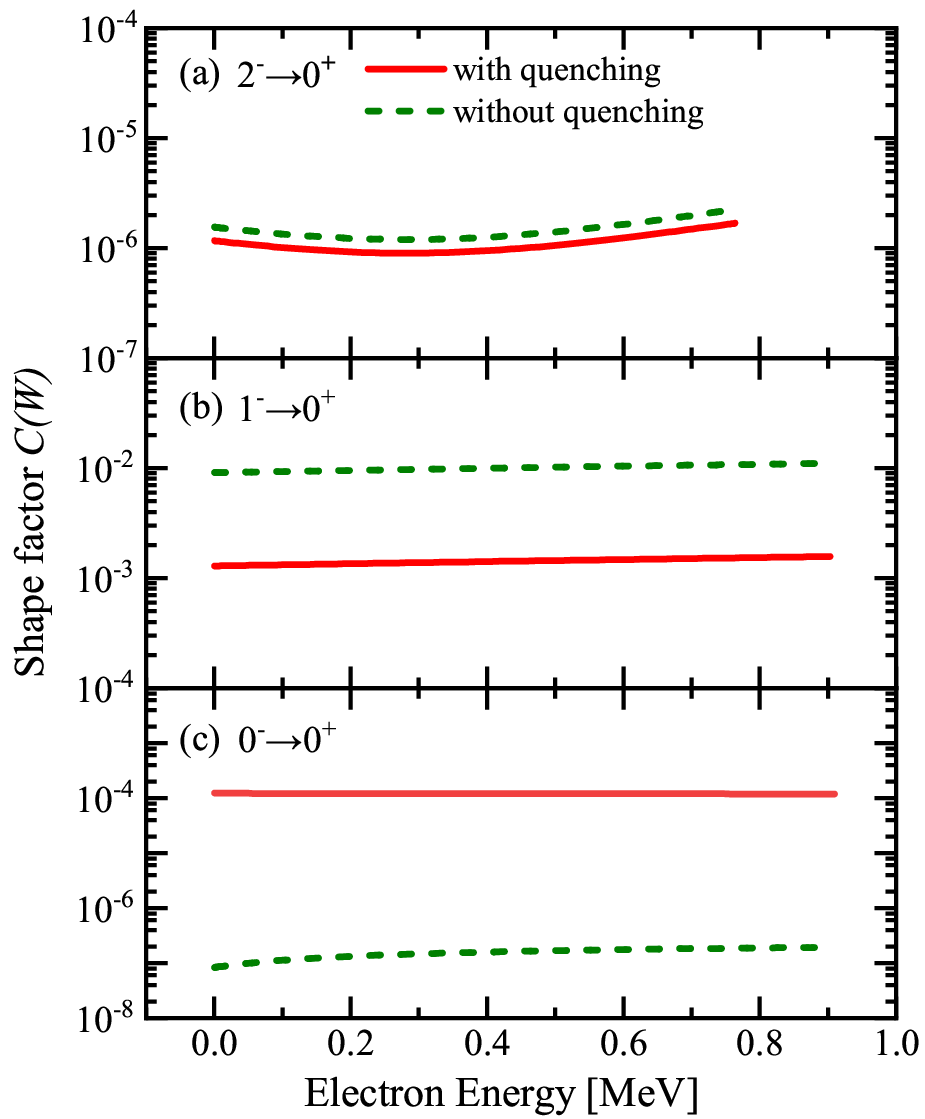}
  \caption{\label{fig:fig5} (color online) The calculated shape factors $C(W)$ for the three first-forbidden transitions of $^{204}$Tl with and without the quenching factors, as a function of the electron kinetic energy $(W-1)m_e c^2$. See the text for details.  }
\end{center}
\end{figure}

\begin{figure}[t]
\begin{center}
  \includegraphics[width=0.49\textwidth]{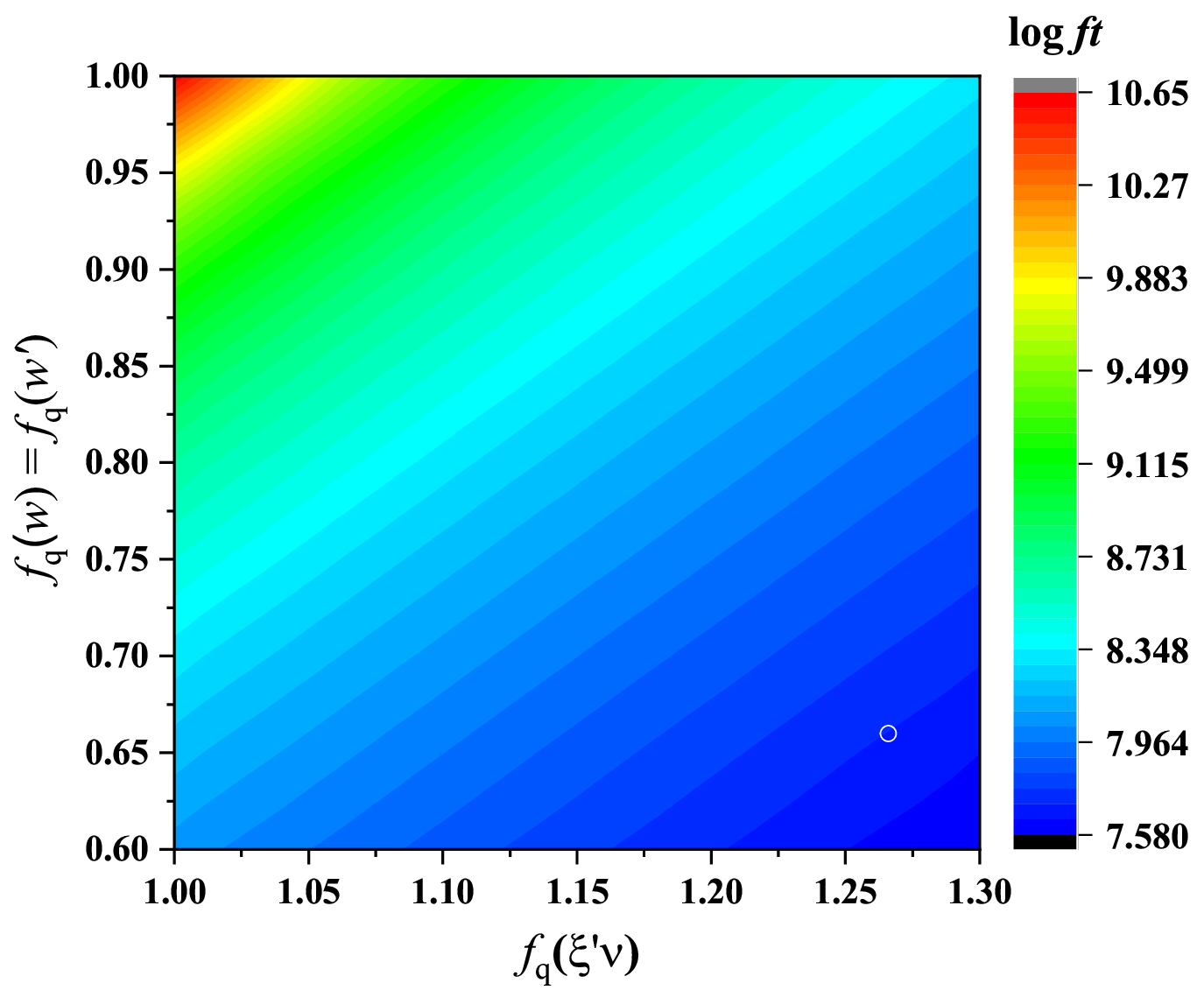}
  \caption{\label{fig:fig6} (color online) The calculated log$ft$ value of the $0^- \rightarrow 0^+$ transition as a function of the corresponding quenching factors discussed in Eq. (\ref{eq.quench}). See the text for details.  }
\end{center}
\end{figure}

It is very interesting as seen from the example of the branching-point $^{204}$Tl that even tiny population probability of nuclear low-lying states would lead to large impact of the stellar $\beta$-decay rates in astrophysical environments with high temperature and high density, provided that the transitions from these low-lying states are predicted to be much stronger than the ones from the corresponding ground states. This indicates that systematical calculations and predictions of stellar $\beta$-decay rates of relevant nuclei, paying special attention to transitions from low-lying states, by elaborate and practical nuclear many-body models considering both allowed and first-forbidden transitions play crucial and indispensable roles in our understanding of the $s$ process, $r$ process etc. The practical PSM model presented in this work provides us one of the candidate numerical models that will be applied systematically to related stellar weak-interaction processes of nuclei in the near future.

As the quenching problem turns out to be important for calculations of transition strengths, the calculated shape factors $C(W)$ for the three first-forbidden transitions of $^{204}$Tl with and without the quenching factors in Eq. (\ref{eq.quench}) are shown in Fig. \ref{fig:fig5} to analyze the effect of quenching. It is seen that when quenching factors are adopted, the shape factor $C(W)$ decreases for the $2^- \rightarrow 0^+$ and $1^- \rightarrow 0^+$ transitions, and increases for the $0^- \rightarrow 0^+$ transition. This indicates that the transition strength (log$ft$ value) decreases (increases) for the $2^- \rightarrow 0^+$ and $1^- \rightarrow 0^+$ transitions, and increases (decreases) for the $0^- \rightarrow 0^+$ transition, when the quenching factors are adopted, which is consistent with Fig. \ref{fig:fig3}. For the $2^- \rightarrow 0^+$ unique first-forbidden transition, only the nuclear matrix element of rank-two tensor operator, i.e., the $z$ term in Eq. (\ref{eq.all_ME}d) contributes and all other terms in Eq. (\ref{eq.all_ME}) vanish. The corresponding quenching factor is adopted as $f_q(z)=0.867$ in Eq. (\ref{eq.quench}) in this work, so that the calculated shape factor $C(W)$ should decrease when such a $f_q(z)$ is adopted, as can be seen from Eqs. (\ref{eq.CW}, \ref{eq.kabc}) and being illustrated in Fig. \ref{fig:fig5}(a). For the $0^- \rightarrow 0^+$ non-unique first-forbidden transition, only the three nuclear matrix elements of scalar operators, i.e., the $w$, $w'$ and $\xi'\nu$ terms in Eq. (\ref{eq.all_ME}) contribute. In this case, the shape factor $C(W)$ is found to be dominated numerically by the $V$ term in Eq. (\ref{eq.VY}). Without adopting the quenching factors, it is found that $\xi'\nu=-1.81\times 10^{-2}$ and $\xi w'=1.73 \times 10^{-2}$ so that they cancel out approximately, which leads to very small value of $V$ that corresponds to small $C(W)$ and large log$ft$ values for the transition, as can be seen from Fig. \ref{fig:fig5}(c) and Fig. \ref{fig:fig6}. When quenching factors $f_q(w/w') < 1.0$ and $f_q(\xi' \nu) > 1.0$ are adopted as in Eq. (\ref{eq.quench}), the approximate cancellation is broken so that $V$ and $C(W)$ increase suddenly with the decrease of $f_q(w')$ and/or the increase of $f_q(\xi' \nu)$. This indicates that the corresponding log$ft$ value could decrease suddenly with the decrease of $f_q(w')$ and/or the increase of $f_q(\xi' \nu)$, as illustrated in Fig. \ref{fig:fig6} where the small circle corresponds to the values of $f_q(w/w')$ and $f_q(\xi' \nu)$ in Eq. (\ref{eq.quench}).

\section{\label{sec:sum}summary and outlook}

In summary, the first-forbidden transitions in nuclear $\beta$ decays are expected to be important for many astrophysical problems regarding stellar nucleosynthesis, such as the $s$ process \cite{TY_table_1987}, $r$ process \cite{Zhi_FF_PRC_2013} etc. In this work we propose a practical model to calculate the stellar $\beta$-decay rates in stellar environments, where contributions from both allowed and first-forbidden transitions are taken into account. The model is based on the recent developments of the projected shell model (PSM) on nuclear first-forbidden transitions in terrestrial case and on Gamow-Teller transitions for stellar weak-interaction rates. An important $s$-process branching-point nucleus, $^{204}$Tl, is taken as the first example, where all related transitions are first-forbidden. In the terrestrial case, the parent nucleus stays in its ground state, the transition from the ground state is unique first-forbidden and is described reasonably by our calculations. With the increase of the $s$-process temperature, although the population probability is tiny, transitions from two low-lying states, which are non-unique first-forbidden, become relevant and are predicted to be stronger than the transition from the ground state by five orders of magnitude from the widely used data tables by Takahashi and Yokoi (TY87), leading to the stellar $\beta$-decay rate of $^{204}$Tl increases very rapidly and making $^{204}$Tl as an $s$-process thermometer and affecting the potential chronometers of the last $s$-process events, $^{205}$Pb. The transitions from two low-lying states are expected to be much reduced from our PSM calculations, and our calculated stellar $\beta$-decay rate of $^{204}$Tl is much lower than the widely used TY87 results by about a factor of five (1.5 orders of magnitude) when no quenching factors (appropriate quenching factors) of nuclear matrix elements are adopted, at the $s$-process typical temperature $T \approx 0.3$ GK. Besides, the effect of the quenching on the decay rate of $^{204}$Tl is found to be as large as about one order of magnitude. 

With the experimental developments of neutron-capture cross sections over the years and in the near future, uncertainty in stellar $\beta$-decay rates need to be reduced with reliable nuclear many-body models for understanding the $s$ process, $r$ process etc. when first-forbidden transition play important roles, the corresponding quenching problem needs further investigation.

\begin{acknowledgments}
  L.J.W. would like to thank B. S. Gao for discussions about charge-exchange reactions and forbidden transitions. This work is supported by the National Natural Science Foundation of China (Grants No. 12275225), and by the Fundamental Research Funds for the Central Universities (Grant No. SWUKT22050). 
\end{acknowledgments}






\begin{thebibliography}{0}%
\makeatletter
\providecommand \@ifxundefined [1]{%
 \@ifx{#1\undefined}
}%
\providecommand \@ifnum [1]{%
 \ifnum #1\expandafter \@firstoftwo
 \else \expandafter \@secondoftwo
 \fi
}%
\providecommand \@ifx [1]{%
 \ifx #1\expandafter \@firstoftwo
 \else \expandafter \@secondoftwo
 \fi
}%
\providecommand \natexlab [1]{#1}%
\providecommand \enquote  [1]{``#1''}%
\providecommand \bibnamefont  [1]{#1}%
\providecommand \bibfnamefont [1]{#1}%
\providecommand \citenamefont [1]{#1}%
\providecommand \href@noop [0]{\@secondoftwo}%
\providecommand \href [0]{\begingroup \@sanitize@url \@href}%
\providecommand \@href[1]{\@@startlink{#1}\@@href}%
\providecommand \@@href[1]{\endgroup#1\@@endlink}%
\providecommand \@sanitize@url [0]{\catcode `\\12\catcode `\$12\catcode
  `\&12\catcode `\#12\catcode `\^12\catcode `\_12\catcode `\%12\relax}%
\providecommand \@@startlink[1]{}%
\providecommand \@@endlink[0]{}%
\providecommand \url  [0]{\begingroup\@sanitize@url \@url }%
\providecommand \@url [1]{\endgroup\@href {#1}{\urlprefix }}%
\providecommand \urlprefix  [0]{URL }%
\providecommand \Eprint [0]{\href }%
\providecommand \doibase [0]{https://doi.org/}%
\providecommand \selectlanguage [0]{\@gobble}%
\providecommand \bibinfo  [0]{\@secondoftwo}%
\providecommand \bibfield  [0]{\@secondoftwo}%
\providecommand \translation [1]{[#1]}%
\providecommand \BibitemOpen [0]{}%
\providecommand \bibitemStop [0]{}%
\providecommand \bibitemNoStop [0]{.\EOS\space}%
\providecommand \EOS [0]{\spacefactor3000\relax}%
\providecommand \BibitemShut  [1]{\csname bibitem#1\endcsname}%
\let\auto@bib@innerbib\@empty
\end{thebibliography}%


\begin{thebibliography}{57}%
\makeatletter
\providecommand \@ifxundefined [1]{%
 \@ifx{#1\undefined}
}%
\providecommand \@ifnum [1]{%
 \ifnum #1\expandafter \@firstoftwo
 \else \expandafter \@secondoftwo
 \fi
}%
\providecommand \@ifx [1]{%
 \ifx #1\expandafter \@firstoftwo
 \else \expandafter \@secondoftwo
 \fi
}%
\providecommand \natexlab [1]{#1}%
\providecommand \enquote  [1]{``#1''}%
\providecommand \bibnamefont  [1]{#1}%
\providecommand \bibfnamefont [1]{#1}%
\providecommand \citenamefont [1]{#1}%
\providecommand \href@noop [0]{\@secondoftwo}%
\providecommand \href [0]{\begingroup \@sanitize@url \@href}%
\providecommand \@href[1]{\@@startlink{#1}\@@href}%
\providecommand \@@href[1]{\endgroup#1\@@endlink}%
\providecommand \@sanitize@url [0]{\catcode `\\12\catcode `\$12\catcode
  `\&12\catcode `\#12\catcode `\^12\catcode `\_12\catcode `\%12\relax}%
\providecommand \@@startlink[1]{}%
\providecommand \@@endlink[0]{}%
\providecommand \url  [0]{\begingroup\@sanitize@url \@url }%
\providecommand \@url [1]{\endgroup\@href {#1}{\urlprefix }}%
\providecommand \urlprefix  [0]{URL }%
\providecommand \Eprint [0]{\href }%
\providecommand \doibase [0]{https://doi.org/}%
\providecommand \selectlanguage [0]{\@gobble}%
\providecommand \bibinfo  [0]{\@secondoftwo}%
\providecommand \bibfield  [0]{\@secondoftwo}%
\providecommand \translation [1]{[#1]}%
\providecommand \BibitemOpen [0]{}%
\providecommand \bibitemStop [0]{}%
\providecommand \bibitemNoStop [0]{.\EOS\space}%
\providecommand \EOS [0]{\spacefactor3000\relax}%
\providecommand \BibitemShut  [1]{\csname bibitem#1\endcsname}%
\let\auto@bib@innerbib\@empty
\bibitem [{\citenamefont {Fuller}\ \emph {et~al.}(1980)\citenamefont {Fuller},
  \citenamefont {Fowler},\ and\ \citenamefont {Newman}}]{Fuller1980}%
  \BibitemOpen
  \bibfield  {author} {\bibinfo {author} {\bibfnamefont {G.~M.}\ \bibnamefont
  {Fuller}}, \bibinfo {author} {\bibfnamefont {W.~A.}\ \bibnamefont {Fowler}},\
  and\ \bibinfo {author} {\bibfnamefont {M.~J.}\ \bibnamefont {Newman}},\
  }\bibfield  {title} {\bibinfo {title} {Stellar weak-interaction rates for
  sd-shell nuclei. i. nuclear matrix element systematics with application to
  $^{26}$\text{Al} and selected nuclei of imprtance to the supernova problem},\
  }\href {https://doi.org/Doi 10.1086/190657} {\bibfield  {journal} {\bibinfo
  {journal} {Astrophys. J. (Suppl.)}\ }\textbf {\bibinfo {volume} {42}},\
  \bibinfo {pages} {447} (\bibinfo {year} {1980})}\BibitemShut {NoStop}%
\bibitem [{\citenamefont {Fuller}\ \emph
  {et~al.}(1982{\natexlab{a}})\citenamefont {Fuller}, \citenamefont {Fowler},\
  and\ \citenamefont {Newman}}]{Fuller1982_1}%
  \BibitemOpen
  \bibfield  {author} {\bibinfo {author} {\bibfnamefont {G.~M.}\ \bibnamefont
  {Fuller}}, \bibinfo {author} {\bibfnamefont {W.~A.}\ \bibnamefont {Fowler}},\
  and\ \bibinfo {author} {\bibfnamefont {M.~J.}\ \bibnamefont {Newman}},\
  }\bibfield  {title} {\bibinfo {title} {Stellar weak-interaction rates for
  intermediate-mass nuclei. ii.},\ }\href@noop {} {\bibfield  {journal}
  {\bibinfo  {journal} {Astrophys. J.}\ }\textbf {\bibinfo {volume} {252}},\
  \bibinfo {pages} {715} (\bibinfo {year} {1982}{\natexlab{a}})}\BibitemShut
  {NoStop}%
\bibitem [{\citenamefont {Fuller}\ \emph
  {et~al.}(1982{\natexlab{b}})\citenamefont {Fuller}, \citenamefont {Fowler},\
  and\ \citenamefont {Newman}}]{Fuller1982_2}%
  \BibitemOpen
  \bibfield  {author} {\bibinfo {author} {\bibfnamefont {G.~M.}\ \bibnamefont
  {Fuller}}, \bibinfo {author} {\bibfnamefont {W.~A.}\ \bibnamefont {Fowler}},\
  and\ \bibinfo {author} {\bibfnamefont {M.~J.}\ \bibnamefont {Newman}},\
  }\bibfield  {title} {\bibinfo {title} {Stellar weak-interaction rates for
  intermediate-mass nuclei. iii.},\ }\href@noop {} {\bibfield  {journal}
  {\bibinfo  {journal} {Astrophys. J. (Suppl.)}\ }\textbf {\bibinfo {volume}
  {48}},\ \bibinfo {pages} {279} (\bibinfo {year}
  {1982}{\natexlab{b}})}\BibitemShut {NoStop}%
\bibitem [{\citenamefont {Fuller}\ \emph {et~al.}(1985)\citenamefont {Fuller},
  \citenamefont {Fowler},\ and\ \citenamefont {Newman}}]{fuller1985}%
  \BibitemOpen
  \bibfield  {author} {\bibinfo {author} {\bibfnamefont {G.~M.}\ \bibnamefont
  {Fuller}}, \bibinfo {author} {\bibfnamefont {W.~A.}\ \bibnamefont {Fowler}},\
  and\ \bibinfo {author} {\bibfnamefont {M.~J.}\ \bibnamefont {Newman}},\
  }\bibfield  {title} {\bibinfo {title} {Stellar weak-interaction rates for
  intermediate-mass nuclei. iv.},\ }\href@noop {} {\bibfield  {journal}
  {\bibinfo  {journal} {Astrophys. J.}\ }\textbf {\bibinfo {volume} {293}},\
  \bibinfo {pages} {1} (\bibinfo {year} {1985})}\BibitemShut {NoStop}%
\bibitem [{\citenamefont {Langanke}\ and\ \citenamefont
  {Mart\'{\i}nez-Pinedo}(2003)}]{langanke_RMP}%
  \BibitemOpen
  \bibfield  {author} {\bibinfo {author} {\bibfnamefont {K.}~\bibnamefont
  {Langanke}}\ and\ \bibinfo {author} {\bibfnamefont {G.}~\bibnamefont
  {Mart\'{\i}nez-Pinedo}},\ }\bibfield  {title} {\bibinfo {title} {Nuclear
  weak-interaction processes in stars},\ }\href
  {https://doi.org/10.1103/RevModPhys.75.819} {\bibfield  {journal} {\bibinfo
  {journal} {Rev. Mod. Phys.}\ }\textbf {\bibinfo {volume} {75}},\ \bibinfo
  {pages} {819} (\bibinfo {year} {2003})}\BibitemShut {NoStop}%
\bibitem [{\citenamefont {Langanke}\ \emph {et~al.}(2021)\citenamefont
  {Langanke}, \citenamefont {Martínez-Pinedo},\ and\ \citenamefont
  {Zegers}}]{langanke_2021_Rep_Pro_Phys}%
  \BibitemOpen
  \bibfield  {author} {\bibinfo {author} {\bibfnamefont {K.}~\bibnamefont
  {Langanke}}, \bibinfo {author} {\bibfnamefont {G.}~\bibnamefont
  {Martínez-Pinedo}},\ and\ \bibinfo {author} {\bibfnamefont {R.~G.~T.}\
  \bibnamefont {Zegers}},\ }\bibfield  {title} {\bibinfo {title} {Electron
  capture in stars},\ }\href {https://doi.org/10.1088/1361-6633/abf207}
  {\bibfield  {journal} {\bibinfo  {journal} {Reports on Progress in Physics}\
  }\textbf {\bibinfo {volume} {84}},\ \bibinfo {pages} {066301} (\bibinfo
  {year} {2021})}\BibitemShut {NoStop}%
\bibitem [{\citenamefont {K\"appeler}\ \emph {et~al.}(2011)\citenamefont
  {K\"appeler}, \citenamefont {Gallino}, \citenamefont {Bisterzo},\ and\
  \citenamefont {Aoki}}]{s_process_RMP_2011}%
  \BibitemOpen
  \bibfield  {author} {\bibinfo {author} {\bibfnamefont {F.}~\bibnamefont
  {K\"appeler}}, \bibinfo {author} {\bibfnamefont {R.}~\bibnamefont {Gallino}},
  \bibinfo {author} {\bibfnamefont {S.}~\bibnamefont {Bisterzo}},\ and\
  \bibinfo {author} {\bibfnamefont {W.}~\bibnamefont {Aoki}},\ }\bibfield
  {title} {\bibinfo {title} {The $s$ process: Nuclear physics, stellar models,
  and observations},\ }\href {https://doi.org/10.1103/RevModPhys.83.157}
  {\bibfield  {journal} {\bibinfo  {journal} {Rev. Mod. Phys.}\ }\textbf
  {\bibinfo {volume} {83}},\ \bibinfo {pages} {157} (\bibinfo {year}
  {2011})}\BibitemShut {NoStop}%
\bibitem [{\citenamefont {Cowan}\ \emph {et~al.}(2021)\citenamefont {Cowan},
  \citenamefont {Sneden}, \citenamefont {Lawler}, \citenamefont {Aprahamian},
  \citenamefont {Wiescher}, \citenamefont {Langanke}, \citenamefont
  {Mart\'{\i}nez-Pinedo},\ and\ \citenamefont
  {Thielemann}}]{r_process_RMP_2021}%
  \BibitemOpen
  \bibfield  {author} {\bibinfo {author} {\bibfnamefont {J.~J.}\ \bibnamefont
  {Cowan}}, \bibinfo {author} {\bibfnamefont {C.}~\bibnamefont {Sneden}},
  \bibinfo {author} {\bibfnamefont {J.~E.}\ \bibnamefont {Lawler}}, \bibinfo
  {author} {\bibfnamefont {A.}~\bibnamefont {Aprahamian}}, \bibinfo {author}
  {\bibfnamefont {M.}~\bibnamefont {Wiescher}}, \bibinfo {author}
  {\bibfnamefont {K.}~\bibnamefont {Langanke}}, \bibinfo {author}
  {\bibfnamefont {G.}~\bibnamefont {Mart\'{\i}nez-Pinedo}},\ and\ \bibinfo
  {author} {\bibfnamefont {F.-K.}\ \bibnamefont {Thielemann}},\ }\bibfield
  {title} {\bibinfo {title} {Origin of the heaviest elements: The rapid
  neutron-capture process},\ }\href
  {https://doi.org/10.1103/RevModPhys.93.015002} {\bibfield  {journal}
  {\bibinfo  {journal} {Rev. Mod. Phys.}\ }\textbf {\bibinfo {volume} {93}},\
  \bibinfo {pages} {015002} (\bibinfo {year} {2021})}\BibitemShut {NoStop}%
\bibitem [{\citenamefont {Schatz}\ \emph {et~al.}(1998)\citenamefont {Schatz},
  \citenamefont {Aprahamian}, \citenamefont {Görres}, \citenamefont
  {Wiescher}, \citenamefont {Rauscher}, \citenamefont {Rembges}, \citenamefont
  {Thielemann}, \citenamefont {Pfeiffer}, \citenamefont {Möller},
  \citenamefont {Kratz}, \citenamefont {Herndl}, \citenamefont {Brown},\ and\
  \citenamefont {Rebel}}]{rp_process_Schatz_1998}%
  \BibitemOpen
  \bibfield  {author} {\bibinfo {author} {\bibfnamefont {H.}~\bibnamefont
  {Schatz}}, \bibinfo {author} {\bibfnamefont {A.}~\bibnamefont {Aprahamian}},
  \bibinfo {author} {\bibfnamefont {J.}~\bibnamefont {Görres}}, \bibinfo
  {author} {\bibfnamefont {M.}~\bibnamefont {Wiescher}}, \bibinfo {author}
  {\bibfnamefont {T.}~\bibnamefont {Rauscher}}, \bibinfo {author}
  {\bibfnamefont {J.}~\bibnamefont {Rembges}}, \bibinfo {author} {\bibfnamefont
  {F.-K.}\ \bibnamefont {Thielemann}}, \bibinfo {author} {\bibfnamefont
  {B.}~\bibnamefont {Pfeiffer}}, \bibinfo {author} {\bibfnamefont
  {P.}~\bibnamefont {Möller}}, \bibinfo {author} {\bibfnamefont {K.-L.}\
  \bibnamefont {Kratz}}, \bibinfo {author} {\bibfnamefont {H.}~\bibnamefont
  {Herndl}}, \bibinfo {author} {\bibfnamefont {B.}~\bibnamefont {Brown}},\ and\
  \bibinfo {author} {\bibfnamefont {H.}~\bibnamefont {Rebel}},\ }\bibfield
  {title} {\bibinfo {title} {rp-process nucleosynthesis at extreme temperature
  and density conditions},\ }\href
  {https://doi.org/https://doi.org/10.1016/S0370-1573(97)00048-3} {\bibfield
  {journal} {\bibinfo  {journal} {Physics Reports}\ }\textbf {\bibinfo {volume}
  {294}},\ \bibinfo {pages} {167} (\bibinfo {year} {1998})}\BibitemShut
  {NoStop}%
\bibitem [{\citenamefont {Schatz}\ \emph {et~al.}(2014)\citenamefont {Schatz},
  \citenamefont {Gupta}, \citenamefont {M\"oller}, \citenamefont {Beard},
  \citenamefont {Brown}, \citenamefont {Deibel}, \citenamefont {Gasques},
  \citenamefont {Hix}, \citenamefont {Keek}, \citenamefont {Lau}, \citenamefont
  {Steiner},\ and\ \citenamefont {Wiescher}}]{schatz2014nature}%
  \BibitemOpen
  \bibfield  {author} {\bibinfo {author} {\bibfnamefont {H.}~\bibnamefont
  {Schatz}}, \bibinfo {author} {\bibfnamefont {S.}~\bibnamefont {Gupta}},
  \bibinfo {author} {\bibfnamefont {P.}~\bibnamefont {M\"oller}}, \bibinfo
  {author} {\bibfnamefont {M.}~\bibnamefont {Beard}}, \bibinfo {author}
  {\bibfnamefont {E.~F.}\ \bibnamefont {Brown}}, \bibinfo {author}
  {\bibfnamefont {A.~T.}\ \bibnamefont {Deibel}}, \bibinfo {author}
  {\bibfnamefont {L.~R.}\ \bibnamefont {Gasques}}, \bibinfo {author}
  {\bibfnamefont {W.~R.}\ \bibnamefont {Hix}}, \bibinfo {author} {\bibfnamefont
  {L.}~\bibnamefont {Keek}}, \bibinfo {author} {\bibfnamefont {R.}~\bibnamefont
  {Lau}}, \bibinfo {author} {\bibfnamefont {A.~W.}\ \bibnamefont {Steiner}},\
  and\ \bibinfo {author} {\bibfnamefont {M.}~\bibnamefont {Wiescher}},\
  }\bibfield  {title} {\bibinfo {title} {Strong neutrino cooling by cycles of
  electron capture and $\beta^-$ decay in neutron star crusts},\ }\href
  {https://doi.org/10.1038/nature12757} {\bibfield  {journal} {\bibinfo
  {journal} {Nature}\ }\textbf {\bibinfo {volume} {505}},\ \bibinfo {pages}
  {62} (\bibinfo {year} {2014})}\BibitemShut {NoStop}%
\bibitem [{\citenamefont {Takahashi}\ and\ \citenamefont
  {Yokoi}(1987)}]{TY_table_1987}%
  \BibitemOpen
  \bibfield  {author} {\bibinfo {author} {\bibfnamefont {K.}~\bibnamefont
  {Takahashi}}\ and\ \bibinfo {author} {\bibfnamefont {K.}~\bibnamefont
  {Yokoi}},\ }\bibfield  {title} {\bibinfo {title} {Beta-decay rates of highly
  ionized heavy atoms in stellar interiors},\ }\href
  {https://doi.org/https://doi.org/10.1016/0092-640X(87)90010-6} {\bibfield
  {journal} {\bibinfo  {journal} {Atomic Data and Nuclear Data Tables}\
  }\textbf {\bibinfo {volume} {36}},\ \bibinfo {pages} {375} (\bibinfo {year}
  {1987})}\BibitemShut {NoStop}%
\bibitem [{\citenamefont {Lugaro}\ \emph {et~al.}(2023)\citenamefont {Lugaro},
  \citenamefont {Pignatari}, \citenamefont {Reifarth},\ and\ \citenamefont
  {Wiescher}}]{Lugaro_Review_2023_s_beyond}%
  \BibitemOpen
  \bibfield  {author} {\bibinfo {author} {\bibfnamefont {M.}~\bibnamefont
  {Lugaro}}, \bibinfo {author} {\bibfnamefont {M.}~\bibnamefont {Pignatari}},
  \bibinfo {author} {\bibfnamefont {R.}~\bibnamefont {Reifarth}},\ and\
  \bibinfo {author} {\bibfnamefont {M.}~\bibnamefont {Wiescher}},\ }\bibfield
  {title} {\bibinfo {title} {The s process and beyond},\ }\href
  {https://doi.org/https://doi.org/10.1146/annurev-nucl-102422-080857}
  {\bibfield  {journal} {\bibinfo  {journal} {Annual Review of Nuclear and
  Particle Science}\ }\textbf {\bibinfo {volume} {73}},\ \bibinfo {pages} {315}
  (\bibinfo {year} {2023})}\BibitemShut {NoStop}%
\bibitem [{\citenamefont {Diehl}(2018)}]{diehl2018astrophysics}%
  \BibitemOpen
  \bibfield  {author} {\bibinfo {author} {\bibfnamefont {R.}~\bibnamefont
  {Diehl}},\ }\href@noop {} {\emph {\bibinfo {title} {Astrophysics with
  radioactive Isotopes}}}\ (\bibinfo  {publisher} {Springer},\ \bibinfo {year}
  {2018})\BibitemShut {NoStop}%
\bibitem [{\citenamefont {Domingo-Pardo}\ \emph {et~al.}(2007)\citenamefont
  {Domingo-Pardo}, \citenamefont {Abbondanno}, \citenamefont {Aerts},
  \citenamefont {\'Alvarez-Pol}, \citenamefont {Alvarez-Velarde}, \citenamefont
  {Andriamonje}, \citenamefont {Andrzejewski}, \citenamefont {Assimakopoulos},
  \citenamefont {Audouin}, \citenamefont {Badurek}, \citenamefont {Baumann},
  \citenamefont {Be\ifmmode \check{c}\else
  \v{c}\fi{}v\'a\ifmmode~\check{r}\else \v{r}\fi{}}, \citenamefont
  {Berthoumieux}, \citenamefont {Bisterzo}, \citenamefont {Calvi\~no},
  \citenamefont {Cano-Ott}, \citenamefont {Capote}, \citenamefont
  {Carrapi\ifmmode~\mbox{\c{c}}\else \c{c}\fi{}o}, \citenamefont {Cennini},
  \citenamefont {Chepel}, \citenamefont {Chiaveri}, \citenamefont {Colonna},
  \citenamefont {Cortes}, \citenamefont {Couture}, \citenamefont {Cox},
  \citenamefont {Dahlfors}, \citenamefont {David}, \citenamefont {Dillmann},
  \citenamefont {Dolfini}, \citenamefont {Dridi}, \citenamefont {Duran},
  \citenamefont {Eleftheriadis}, \citenamefont {Embid-Segura}, \citenamefont
  {Ferrant}, \citenamefont {Ferrari}, \citenamefont {Ferreira-Marques},
  \citenamefont {Fitzpatrick}, \citenamefont {Frais-Koelbl}, \citenamefont
  {Fujii}, \citenamefont {Furman}, \citenamefont {Gallino}, \citenamefont
  {Goncalves}, \citenamefont {Gonzalez-Romero}, \citenamefont {Goverdovski},
  \citenamefont {Gramegna}, \citenamefont {Griesmayer}, \citenamefont
  {Guerrero}, \citenamefont {Gunsing}, \citenamefont {Haas}, \citenamefont
  {Haight}, \citenamefont {Heil}, \citenamefont {Herrera-Martinez},
  \citenamefont {Igashira}, \citenamefont {Isaev}, \citenamefont {Jericha},
  \citenamefont {Kadi}, \citenamefont {K\"appeler}, \citenamefont {Karamanis},
  \citenamefont {Karadimos}, \citenamefont {Kerveno}, \citenamefont {Ketlerov},
  \citenamefont {Koehler}, \citenamefont {Konovalov}, \citenamefont
  {Kossionides}, \citenamefont {Krti\ifmmode~\check{c}\else \v{c}\fi{}ka},
  \citenamefont {Lamboudis}, \citenamefont {Leeb}, \citenamefont {Lindote},
  \citenamefont {Lopes}, \citenamefont {Lozano}, \citenamefont {Lukic},
  \citenamefont {Marganiec}, \citenamefont {Marrone}, \citenamefont {Mastinu},
  \citenamefont {Mengoni}, \citenamefont {Milazzo}, \citenamefont {Moreau},
  \citenamefont {Mosconi}, \citenamefont {Neves}, \citenamefont {Oberhummer},
  \citenamefont {Oshima}, \citenamefont {O'Brien}, \citenamefont {Pancin},
  \citenamefont {Papachristodoulou}, \citenamefont {Papadopoulos},
  \citenamefont {Paradela}, \citenamefont {Patronis}, \citenamefont {Pavlik},
  \citenamefont {Pavlopoulos}, \citenamefont {Perrot}, \citenamefont {Plag},
  \citenamefont {Plompen}, \citenamefont {Plukis}, \citenamefont {Poch},
  \citenamefont {Pretel}, \citenamefont {Quesada}, \citenamefont {Rauscher},
  \citenamefont {Reifarth}, \citenamefont {Rosetti}, \citenamefont {Rubbia},
  \citenamefont {Rudolf}, \citenamefont {Rullhusen}, \citenamefont {Salgado},
  \citenamefont {Sarchiapone}, \citenamefont {Savvidis}, \citenamefont
  {Stephan}, \citenamefont {Tagliente}, \citenamefont {Tain}, \citenamefont
  {Tassan-Got}, \citenamefont {Tavora}, \citenamefont {Terlizzi}, \citenamefont
  {Vannini}, \citenamefont {Vaz}, \citenamefont {Ventura}, \citenamefont
  {Villamarin}, \citenamefont {Vincente}, \citenamefont {Vlachoudis},
  \citenamefont {Vlastou}, \citenamefont {Voss}, \citenamefont {Walter},
  \citenamefont {Wendler}, \citenamefont {Wiescher},\ and\ \citenamefont
  {Wisshak}}]{Doming0_PRC_2007}%
  \BibitemOpen
  \bibfield  {author} {\bibinfo {author} {\bibfnamefont {C.}~\bibnamefont
  {Domingo-Pardo}}, \bibinfo {author} {\bibfnamefont {U.}~\bibnamefont
  {Abbondanno}}, \bibinfo {author} {\bibfnamefont {G.}~\bibnamefont {Aerts}},
  \bibinfo {author} {\bibfnamefont {H.}~\bibnamefont {\'Alvarez-Pol}}, \bibinfo
  {author} {\bibfnamefont {F.}~\bibnamefont {Alvarez-Velarde}}, \bibinfo
  {author} {\bibfnamefont {S.}~\bibnamefont {Andriamonje}}, \bibinfo {author}
  {\bibfnamefont {J.}~\bibnamefont {Andrzejewski}}, \bibinfo {author}
  {\bibfnamefont {P.}~\bibnamefont {Assimakopoulos}}, \bibinfo {author}
  {\bibfnamefont {L.}~\bibnamefont {Audouin}}, \bibinfo {author} {\bibfnamefont
  {G.}~\bibnamefont {Badurek}}, \bibinfo {author} {\bibfnamefont
  {P.}~\bibnamefont {Baumann}}, \bibinfo {author} {\bibfnamefont
  {F.}~\bibnamefont {Be\ifmmode \check{c}\else
  \v{c}\fi{}v\'a\ifmmode~\check{r}\else \v{r}\fi{}}}, \bibinfo {author}
  {\bibfnamefont {E.}~\bibnamefont {Berthoumieux}}, \bibinfo {author}
  {\bibfnamefont {S.}~\bibnamefont {Bisterzo}}, \bibinfo {author}
  {\bibfnamefont {F.}~\bibnamefont {Calvi\~no}}, \bibinfo {author}
  {\bibfnamefont {D.}~\bibnamefont {Cano-Ott}}, \bibinfo {author}
  {\bibfnamefont {R.}~\bibnamefont {Capote}}, \bibinfo {author} {\bibfnamefont
  {C.}~\bibnamefont {Carrapi\ifmmode~\mbox{\c{c}}\else \c{c}\fi{}o}}, \bibinfo
  {author} {\bibfnamefont {P.}~\bibnamefont {Cennini}}, \bibinfo {author}
  {\bibfnamefont {V.}~\bibnamefont {Chepel}}, \bibinfo {author} {\bibfnamefont
  {E.}~\bibnamefont {Chiaveri}}, \bibinfo {author} {\bibfnamefont
  {N.}~\bibnamefont {Colonna}}, \bibinfo {author} {\bibfnamefont
  {G.}~\bibnamefont {Cortes}}, \bibinfo {author} {\bibfnamefont
  {A.}~\bibnamefont {Couture}}, \bibinfo {author} {\bibfnamefont
  {J.}~\bibnamefont {Cox}}, \bibinfo {author} {\bibfnamefont {M.}~\bibnamefont
  {Dahlfors}}, \bibinfo {author} {\bibfnamefont {S.}~\bibnamefont {David}},
  \bibinfo {author} {\bibfnamefont {I.}~\bibnamefont {Dillmann}}, \bibinfo
  {author} {\bibfnamefont {R.}~\bibnamefont {Dolfini}}, \bibinfo {author}
  {\bibfnamefont {W.}~\bibnamefont {Dridi}}, \bibinfo {author} {\bibfnamefont
  {I.}~\bibnamefont {Duran}}, \bibinfo {author} {\bibfnamefont
  {C.}~\bibnamefont {Eleftheriadis}}, \bibinfo {author} {\bibfnamefont
  {M.}~\bibnamefont {Embid-Segura}}, \bibinfo {author} {\bibfnamefont
  {L.}~\bibnamefont {Ferrant}}, \bibinfo {author} {\bibfnamefont
  {A.}~\bibnamefont {Ferrari}}, \bibinfo {author} {\bibfnamefont
  {R.}~\bibnamefont {Ferreira-Marques}}, \bibinfo {author} {\bibfnamefont
  {L.}~\bibnamefont {Fitzpatrick}}, \bibinfo {author} {\bibfnamefont
  {H.}~\bibnamefont {Frais-Koelbl}}, \bibinfo {author} {\bibfnamefont
  {K.}~\bibnamefont {Fujii}}, \bibinfo {author} {\bibfnamefont
  {W.}~\bibnamefont {Furman}}, \bibinfo {author} {\bibfnamefont
  {R.}~\bibnamefont {Gallino}}, \bibinfo {author} {\bibfnamefont
  {I.}~\bibnamefont {Goncalves}}, \bibinfo {author} {\bibfnamefont
  {E.}~\bibnamefont {Gonzalez-Romero}}, \bibinfo {author} {\bibfnamefont
  {A.}~\bibnamefont {Goverdovski}}, \bibinfo {author} {\bibfnamefont
  {F.}~\bibnamefont {Gramegna}}, \bibinfo {author} {\bibfnamefont
  {E.}~\bibnamefont {Griesmayer}}, \bibinfo {author} {\bibfnamefont
  {C.}~\bibnamefont {Guerrero}}, \bibinfo {author} {\bibfnamefont
  {F.}~\bibnamefont {Gunsing}}, \bibinfo {author} {\bibfnamefont
  {B.}~\bibnamefont {Haas}}, \bibinfo {author} {\bibfnamefont {R.}~\bibnamefont
  {Haight}}, \bibinfo {author} {\bibfnamefont {M.}~\bibnamefont {Heil}},
  \bibinfo {author} {\bibfnamefont {A.}~\bibnamefont {Herrera-Martinez}},
  \bibinfo {author} {\bibfnamefont {M.}~\bibnamefont {Igashira}}, \bibinfo
  {author} {\bibfnamefont {S.}~\bibnamefont {Isaev}}, \bibinfo {author}
  {\bibfnamefont {E.}~\bibnamefont {Jericha}}, \bibinfo {author} {\bibfnamefont
  {Y.}~\bibnamefont {Kadi}}, \bibinfo {author} {\bibfnamefont {F.}~\bibnamefont
  {K\"appeler}}, \bibinfo {author} {\bibfnamefont {D.}~\bibnamefont
  {Karamanis}}, \bibinfo {author} {\bibfnamefont {D.}~\bibnamefont
  {Karadimos}}, \bibinfo {author} {\bibfnamefont {M.}~\bibnamefont {Kerveno}},
  \bibinfo {author} {\bibfnamefont {V.}~\bibnamefont {Ketlerov}}, \bibinfo
  {author} {\bibfnamefont {P.}~\bibnamefont {Koehler}}, \bibinfo {author}
  {\bibfnamefont {V.}~\bibnamefont {Konovalov}}, \bibinfo {author}
  {\bibfnamefont {E.}~\bibnamefont {Kossionides}}, \bibinfo {author}
  {\bibfnamefont {M.}~\bibnamefont {Krti\ifmmode~\check{c}\else \v{c}\fi{}ka}},
  \bibinfo {author} {\bibfnamefont {C.}~\bibnamefont {Lamboudis}}, \bibinfo
  {author} {\bibfnamefont {H.}~\bibnamefont {Leeb}}, \bibinfo {author}
  {\bibfnamefont {A.}~\bibnamefont {Lindote}}, \bibinfo {author} {\bibfnamefont
  {I.}~\bibnamefont {Lopes}}, \bibinfo {author} {\bibfnamefont
  {M.}~\bibnamefont {Lozano}}, \bibinfo {author} {\bibfnamefont
  {S.}~\bibnamefont {Lukic}}, \bibinfo {author} {\bibfnamefont
  {J.}~\bibnamefont {Marganiec}}, \bibinfo {author} {\bibfnamefont
  {S.}~\bibnamefont {Marrone}}, \bibinfo {author} {\bibfnamefont
  {P.}~\bibnamefont {Mastinu}}, \bibinfo {author} {\bibfnamefont
  {A.}~\bibnamefont {Mengoni}}, \bibinfo {author} {\bibfnamefont {P.~M.}\
  \bibnamefont {Milazzo}}, \bibinfo {author} {\bibfnamefont {C.}~\bibnamefont
  {Moreau}}, \bibinfo {author} {\bibfnamefont {M.}~\bibnamefont {Mosconi}},
  \bibinfo {author} {\bibfnamefont {F.}~\bibnamefont {Neves}}, \bibinfo
  {author} {\bibfnamefont {H.}~\bibnamefont {Oberhummer}}, \bibinfo {author}
  {\bibfnamefont {M.}~\bibnamefont {Oshima}}, \bibinfo {author} {\bibfnamefont
  {S.}~\bibnamefont {O'Brien}}, \bibinfo {author} {\bibfnamefont
  {J.}~\bibnamefont {Pancin}}, \bibinfo {author} {\bibfnamefont
  {C.}~\bibnamefont {Papachristodoulou}}, \bibinfo {author} {\bibfnamefont
  {C.}~\bibnamefont {Papadopoulos}}, \bibinfo {author} {\bibfnamefont
  {C.}~\bibnamefont {Paradela}}, \bibinfo {author} {\bibfnamefont
  {N.}~\bibnamefont {Patronis}}, \bibinfo {author} {\bibfnamefont
  {A.}~\bibnamefont {Pavlik}}, \bibinfo {author} {\bibfnamefont
  {P.}~\bibnamefont {Pavlopoulos}}, \bibinfo {author} {\bibfnamefont
  {L.}~\bibnamefont {Perrot}}, \bibinfo {author} {\bibfnamefont
  {R.}~\bibnamefont {Plag}}, \bibinfo {author} {\bibfnamefont {A.}~\bibnamefont
  {Plompen}}, \bibinfo {author} {\bibfnamefont {A.}~\bibnamefont {Plukis}},
  \bibinfo {author} {\bibfnamefont {A.}~\bibnamefont {Poch}}, \bibinfo {author}
  {\bibfnamefont {C.}~\bibnamefont {Pretel}}, \bibinfo {author} {\bibfnamefont
  {J.}~\bibnamefont {Quesada}}, \bibinfo {author} {\bibfnamefont
  {T.}~\bibnamefont {Rauscher}}, \bibinfo {author} {\bibfnamefont
  {R.}~\bibnamefont {Reifarth}}, \bibinfo {author} {\bibfnamefont
  {M.}~\bibnamefont {Rosetti}}, \bibinfo {author} {\bibfnamefont
  {C.}~\bibnamefont {Rubbia}}, \bibinfo {author} {\bibfnamefont
  {G.}~\bibnamefont {Rudolf}}, \bibinfo {author} {\bibfnamefont
  {P.}~\bibnamefont {Rullhusen}}, \bibinfo {author} {\bibfnamefont
  {J.}~\bibnamefont {Salgado}}, \bibinfo {author} {\bibfnamefont
  {L.}~\bibnamefont {Sarchiapone}}, \bibinfo {author} {\bibfnamefont
  {I.}~\bibnamefont {Savvidis}}, \bibinfo {author} {\bibfnamefont
  {C.}~\bibnamefont {Stephan}}, \bibinfo {author} {\bibfnamefont
  {G.}~\bibnamefont {Tagliente}}, \bibinfo {author} {\bibfnamefont {J.~L.}\
  \bibnamefont {Tain}}, \bibinfo {author} {\bibfnamefont {L.}~\bibnamefont
  {Tassan-Got}}, \bibinfo {author} {\bibfnamefont {L.}~\bibnamefont {Tavora}},
  \bibinfo {author} {\bibfnamefont {R.}~\bibnamefont {Terlizzi}}, \bibinfo
  {author} {\bibfnamefont {G.}~\bibnamefont {Vannini}}, \bibinfo {author}
  {\bibfnamefont {P.}~\bibnamefont {Vaz}}, \bibinfo {author} {\bibfnamefont
  {A.}~\bibnamefont {Ventura}}, \bibinfo {author} {\bibfnamefont
  {D.}~\bibnamefont {Villamarin}}, \bibinfo {author} {\bibfnamefont {M.~C.}\
  \bibnamefont {Vincente}}, \bibinfo {author} {\bibfnamefont {V.}~\bibnamefont
  {Vlachoudis}}, \bibinfo {author} {\bibfnamefont {R.}~\bibnamefont {Vlastou}},
  \bibinfo {author} {\bibfnamefont {F.}~\bibnamefont {Voss}}, \bibinfo {author}
  {\bibfnamefont {S.}~\bibnamefont {Walter}}, \bibinfo {author} {\bibfnamefont
  {H.}~\bibnamefont {Wendler}}, \bibinfo {author} {\bibfnamefont
  {M.}~\bibnamefont {Wiescher}},\ and\ \bibinfo {author} {\bibfnamefont
  {K.}~\bibnamefont {Wisshak}},\ }\bibfield  {title} {\bibinfo {title}
  {Measurement of the neutron capture cross section of the $s$-only isotope
  $^{204}\mathrm{Pb}$ from 1 ev to 440 kev},\ }\href
  {https://doi.org/10.1103/PhysRevC.75.015806} {\bibfield  {journal} {\bibinfo
  {journal} {Phys. Rev. C}\ }\textbf {\bibinfo {volume} {75}},\ \bibinfo
  {pages} {015806} (\bibinfo {year} {2007})}\BibitemShut {NoStop}%
\bibitem [{\citenamefont {Ratzel}\ \emph {et~al.}(2004)\citenamefont {Ratzel},
  \citenamefont {Arlandini}, \citenamefont {K\"appeler}, \citenamefont
  {Couture}, \citenamefont {Wiescher}, \citenamefont {Reifarth}, \citenamefont
  {Gallino}, \citenamefont {Mengoni},\ and\ \citenamefont
  {Travaglio}}]{Ratzel_PRC_2004}%
  \BibitemOpen
  \bibfield  {author} {\bibinfo {author} {\bibfnamefont {U.}~\bibnamefont
  {Ratzel}}, \bibinfo {author} {\bibfnamefont {C.}~\bibnamefont {Arlandini}},
  \bibinfo {author} {\bibfnamefont {F.}~\bibnamefont {K\"appeler}}, \bibinfo
  {author} {\bibfnamefont {A.}~\bibnamefont {Couture}}, \bibinfo {author}
  {\bibfnamefont {M.}~\bibnamefont {Wiescher}}, \bibinfo {author}
  {\bibfnamefont {R.}~\bibnamefont {Reifarth}}, \bibinfo {author}
  {\bibfnamefont {R.}~\bibnamefont {Gallino}}, \bibinfo {author} {\bibfnamefont
  {A.}~\bibnamefont {Mengoni}},\ and\ \bibinfo {author} {\bibfnamefont
  {C.}~\bibnamefont {Travaglio}},\ }\bibfield  {title} {\bibinfo {title}
  {Nucleosynthesis at the termination point of the $s$ process},\ }\href
  {https://doi.org/10.1103/PhysRevC.70.065803} {\bibfield  {journal} {\bibinfo
  {journal} {Phys. Rev. C}\ }\textbf {\bibinfo {volume} {70}},\ \bibinfo
  {pages} {065803} (\bibinfo {year} {2004})}\BibitemShut {NoStop}%
\bibitem [{\citenamefont {Broda}\ \emph {et~al.}(2011)\citenamefont {Broda},
  \citenamefont {Maier}, \citenamefont {Fornal}, \citenamefont
  {Wrzesi\ifmmode~\acute{n}\else \'{n}\fi{}ski}, \citenamefont {Szpak},
  \citenamefont {Carpenter}, \citenamefont {Janssens}, \citenamefont
  {Kr\'olas}, \citenamefont {Paw\l{}at},\ and\ \citenamefont
  {Zhu}}]{Broda_PRC_2011_high_spin}%
  \BibitemOpen
  \bibfield  {author} {\bibinfo {author} {\bibfnamefont {R.}~\bibnamefont
  {Broda}}, \bibinfo {author} {\bibfnamefont {K.~H.}\ \bibnamefont {Maier}},
  \bibinfo {author} {\bibfnamefont {B.}~\bibnamefont {Fornal}}, \bibinfo
  {author} {\bibfnamefont {J.}~\bibnamefont {Wrzesi\ifmmode~\acute{n}\else
  \'{n}\fi{}ski}}, \bibinfo {author} {\bibfnamefont {B.}~\bibnamefont {Szpak}},
  \bibinfo {author} {\bibfnamefont {M.~P.}\ \bibnamefont {Carpenter}}, \bibinfo
  {author} {\bibfnamefont {R.~V.~F.}\ \bibnamefont {Janssens}}, \bibinfo
  {author} {\bibfnamefont {W.}~\bibnamefont {Kr\'olas}}, \bibinfo {author}
  {\bibfnamefont {T.}~\bibnamefont {Paw\l{}at}},\ and\ \bibinfo {author}
  {\bibfnamefont {S.}~\bibnamefont {Zhu}},\ }\bibfield  {title} {\bibinfo
  {title} {High-spin states and isomers in the one-proton-hole and
  three-neutron-hole ${}^{204}$tl isotope},\ }\href
  {https://doi.org/10.1103/PhysRevC.84.014330} {\bibfield  {journal} {\bibinfo
  {journal} {Phys. Rev. C}\ }\textbf {\bibinfo {volume} {84}},\ \bibinfo
  {pages} {014330} (\bibinfo {year} {2011})}\BibitemShut {NoStop}%
\bibitem [{\citenamefont {Utsunomiya}\ \emph {et~al.}(2019)\citenamefont
  {Utsunomiya}, \citenamefont {Renstr\o{}m}, \citenamefont {Tveten},
  \citenamefont {Goriely}, \citenamefont {Ari-izumi}, \citenamefont
  {Filipescu}, \citenamefont {Kaur}, \citenamefont {Lui}, \citenamefont {Luo},
  \citenamefont {Miyamoto}, \citenamefont {Larsen}, \citenamefont {Hilaire},
  \citenamefont {P\'eru},\ and\ \citenamefont {Koning}}]{Utsunomiya_PRC_2019}%
  \BibitemOpen
  \bibfield  {author} {\bibinfo {author} {\bibfnamefont {H.}~\bibnamefont
  {Utsunomiya}}, \bibinfo {author} {\bibfnamefont {T.}~\bibnamefont
  {Renstr\o{}m}}, \bibinfo {author} {\bibfnamefont {G.~M.}\ \bibnamefont
  {Tveten}}, \bibinfo {author} {\bibfnamefont {S.}~\bibnamefont {Goriely}},
  \bibinfo {author} {\bibfnamefont {T.}~\bibnamefont {Ari-izumi}}, \bibinfo
  {author} {\bibfnamefont {D.}~\bibnamefont {Filipescu}}, \bibinfo {author}
  {\bibfnamefont {J.}~\bibnamefont {Kaur}}, \bibinfo {author} {\bibfnamefont
  {Y.-W.}\ \bibnamefont {Lui}}, \bibinfo {author} {\bibfnamefont
  {W.}~\bibnamefont {Luo}}, \bibinfo {author} {\bibfnamefont {S.}~\bibnamefont
  {Miyamoto}}, \bibinfo {author} {\bibfnamefont {A.~C.}\ \bibnamefont
  {Larsen}}, \bibinfo {author} {\bibfnamefont {S.}~\bibnamefont {Hilaire}},
  \bibinfo {author} {\bibfnamefont {S.}~\bibnamefont {P\'eru}},\ and\ \bibinfo
  {author} {\bibfnamefont {A.~J.}\ \bibnamefont {Koning}},\ }\bibfield  {title}
  {\bibinfo {title} {$\ensuremath{\gamma}$-ray strength function for thallium
  isotopes relevant to the $^{205}\mathrm{Pb}\ensuremath{-}^{205}\mathrm{Tl}$
  chronometry},\ }\href {https://doi.org/10.1103/PhysRevC.99.024609} {\bibfield
   {journal} {\bibinfo  {journal} {Phys. Rev. C}\ }\textbf {\bibinfo {volume}
  {99}},\ \bibinfo {pages} {024609} (\bibinfo {year} {2019})}\BibitemShut
  {NoStop}%
\bibitem [{\citenamefont {Casanovas}\ \emph {et~al.}(2020)\citenamefont
  {Casanovas}, \citenamefont {Tarifeño-Saldivia}, \citenamefont
  {Domingo-Pardo} \emph {et~al.}}]{Casanovas_2020_JP_Con}%
  \BibitemOpen
  \bibfield  {author} {\bibinfo {author} {\bibfnamefont {A.}~\bibnamefont
  {Casanovas}}, \bibinfo {author} {\bibfnamefont {A.~E.}\ \bibnamefont
  {Tarifeño-Saldivia}}, \bibinfo {author} {\bibfnamefont {C.}~\bibnamefont
  {Domingo-Pardo}}, \emph {et~al.},\ }\bibfield  {title} {\bibinfo {title}
  {Neutron capture measurement at the n tof facility of the $^{204}$tl and
  $^{205}$tl s-process branching points},\ }\href
  {https://doi.org/10.1088/1742-6596/1668/1/012005} {\bibfield  {journal}
  {\bibinfo  {journal} {Journal of Physics: Conference Series}\ }\textbf
  {\bibinfo {volume} {1668}},\ \bibinfo {pages} {012005} (\bibinfo {year}
  {2020})}\BibitemShut {NoStop}%
\bibitem [{\citenamefont {A.~Casanovas}\ \emph {et~al.}(2018)\citenamefont
  {A.~Casanovas}, \citenamefont {Domingo-Pardo}, \citenamefont {Guerrero} \emph
  {et~al.}}]{Casanovas_2018_EPJ_Web}%
  \BibitemOpen
  \bibfield  {author} {\bibinfo {author} {\bibfnamefont {A.}~\bibnamefont
  {A.~Casanovas}}, \bibinfo {author} {\bibfnamefont {C.}~\bibnamefont
  {Domingo-Pardo}}, \bibinfo {author} {\bibfnamefont {C.}~\bibnamefont
  {Guerrero}}, \emph {et~al.},\ }\bibfield  {title} {\bibinfo {title}
  {Measurement of the radiative capture cross section of the s-process
  branching points $^{204}$tl and $^{171}$tm at the n tof facility (cern)},\
  }\href {https://doi.org/10.1051/epjconf/201817803004} {\bibfield  {journal}
  {\bibinfo  {journal} {EPJ Web of Conferences}\ }\textbf {\bibinfo {volume}
  {178}},\ \bibinfo {pages} {03004} (\bibinfo {year} {2018})}\BibitemShut
  {NoStop}%
\bibitem [{\citenamefont {Casanovas-Hoste}\ \emph {et~al.}(2022)\citenamefont
  {Casanovas-Hoste}, \citenamefont {Domingo-Pardo}, \citenamefont {Guerrero}
  \emph {et~al.}}]{Casanovas_2022_EPJ_Web}%
  \BibitemOpen
  \bibfield  {author} {\bibinfo {author} {\bibfnamefont {A.}~\bibnamefont
  {Casanovas-Hoste}}, \bibinfo {author} {\bibfnamefont {C.}~\bibnamefont
  {Domingo-Pardo}}, \bibinfo {author} {\bibfnamefont {C.}~\bibnamefont
  {Guerrero}}, \emph {et~al.},\ }\bibfield  {title} {\bibinfo {title} {Analysis
  of the impact of the $^{204}$tl neutron capture cross section on the
  s-process only isotope $^{204}$pb},\ }\href
  {https://doi.org/10.1051/epjconf/202226002002} {\bibfield  {journal}
  {\bibinfo  {journal} {EPJ Web of Conferences}\ }\textbf {\bibinfo {volume}
  {260}},\ \bibinfo {pages} {02002} (\bibinfo {year} {2022})}\BibitemShut
  {NoStop}%
\bibitem [{\citenamefont {Gao}\ \emph {et~al.}(2021)\citenamefont {Gao},
  \citenamefont {Giraud}, \citenamefont {Li}, \citenamefont {Sieverding},
  \citenamefont {Zegers}, \citenamefont {Tang}, \citenamefont {Ash},
  \citenamefont {Ayyad-Limonge}, \citenamefont {Bazin}, \citenamefont {Biswas},
  \citenamefont {Brown}, \citenamefont {Chen}, \citenamefont {DeNudt},
  \citenamefont {Farris}, \citenamefont {Gabler}, \citenamefont {Gade},
  \citenamefont {Ginter}, \citenamefont {Grinder}, \citenamefont {Heger},
  \citenamefont {Hultquist}, \citenamefont {Hill}, \citenamefont {Iwasaki},
  \citenamefont {Kwan}, \citenamefont {Li}, \citenamefont {Longfellow},
  \citenamefont {Maher}, \citenamefont {Ndayisabye}, \citenamefont {Noji},
  \citenamefont {Pereira}, \citenamefont {Qi}, \citenamefont {Rebenstock},
  \citenamefont {Revel}, \citenamefont {Rhodes}, \citenamefont {Sanchez},
  \citenamefont {Schmitt}, \citenamefont {Sumithrarachchi}, \citenamefont
  {Sun},\ and\ \citenamefont {Weisshaar}}]{BGao_2021_PRL_59Fe}%
  \BibitemOpen
  \bibfield  {author} {\bibinfo {author} {\bibfnamefont {B.}~\bibnamefont
  {Gao}}, \bibinfo {author} {\bibfnamefont {S.}~\bibnamefont {Giraud}},
  \bibinfo {author} {\bibfnamefont {K.~A.}\ \bibnamefont {Li}}, \bibinfo
  {author} {\bibfnamefont {A.}~\bibnamefont {Sieverding}}, \bibinfo {author}
  {\bibfnamefont {R.~G.~T.}\ \bibnamefont {Zegers}}, \bibinfo {author}
  {\bibfnamefont {X.}~\bibnamefont {Tang}}, \bibinfo {author} {\bibfnamefont
  {J.}~\bibnamefont {Ash}}, \bibinfo {author} {\bibfnamefont {Y.}~\bibnamefont
  {Ayyad-Limonge}}, \bibinfo {author} {\bibfnamefont {D.}~\bibnamefont
  {Bazin}}, \bibinfo {author} {\bibfnamefont {S.}~\bibnamefont {Biswas}},
  \bibinfo {author} {\bibfnamefont {B.~A.}\ \bibnamefont {Brown}}, \bibinfo
  {author} {\bibfnamefont {J.}~\bibnamefont {Chen}}, \bibinfo {author}
  {\bibfnamefont {M.}~\bibnamefont {DeNudt}}, \bibinfo {author} {\bibfnamefont
  {P.}~\bibnamefont {Farris}}, \bibinfo {author} {\bibfnamefont {J.~M.}\
  \bibnamefont {Gabler}}, \bibinfo {author} {\bibfnamefont {A.}~\bibnamefont
  {Gade}}, \bibinfo {author} {\bibfnamefont {T.}~\bibnamefont {Ginter}},
  \bibinfo {author} {\bibfnamefont {M.}~\bibnamefont {Grinder}}, \bibinfo
  {author} {\bibfnamefont {A.}~\bibnamefont {Heger}}, \bibinfo {author}
  {\bibfnamefont {C.}~\bibnamefont {Hultquist}}, \bibinfo {author}
  {\bibfnamefont {A.~M.}\ \bibnamefont {Hill}}, \bibinfo {author}
  {\bibfnamefont {H.}~\bibnamefont {Iwasaki}}, \bibinfo {author} {\bibfnamefont
  {E.}~\bibnamefont {Kwan}}, \bibinfo {author} {\bibfnamefont {J.}~\bibnamefont
  {Li}}, \bibinfo {author} {\bibfnamefont {B.}~\bibnamefont {Longfellow}},
  \bibinfo {author} {\bibfnamefont {C.}~\bibnamefont {Maher}}, \bibinfo
  {author} {\bibfnamefont {F.}~\bibnamefont {Ndayisabye}}, \bibinfo {author}
  {\bibfnamefont {S.}~\bibnamefont {Noji}}, \bibinfo {author} {\bibfnamefont
  {J.}~\bibnamefont {Pereira}}, \bibinfo {author} {\bibfnamefont
  {C.}~\bibnamefont {Qi}}, \bibinfo {author} {\bibfnamefont {J.}~\bibnamefont
  {Rebenstock}}, \bibinfo {author} {\bibfnamefont {A.}~\bibnamefont {Revel}},
  \bibinfo {author} {\bibfnamefont {D.}~\bibnamefont {Rhodes}}, \bibinfo
  {author} {\bibfnamefont {A.}~\bibnamefont {Sanchez}}, \bibinfo {author}
  {\bibfnamefont {J.}~\bibnamefont {Schmitt}}, \bibinfo {author} {\bibfnamefont
  {C.}~\bibnamefont {Sumithrarachchi}}, \bibinfo {author} {\bibfnamefont
  {B.~H.}\ \bibnamefont {Sun}},\ and\ \bibinfo {author} {\bibfnamefont
  {D.}~\bibnamefont {Weisshaar}},\ }\bibfield  {title} {\bibinfo {title} {New
  $^{59}\mathrm{Fe}$ stellar decay rate with implications for the
  $^{60}\mathrm{Fe}$ radioactivity in massive stars},\ }\href
  {https://doi.org/10.1103/PhysRevLett.126.152701} {\bibfield  {journal}
  {\bibinfo  {journal} {Phys. Rev. Lett.}\ }\textbf {\bibinfo {volume} {126}},\
  \bibinfo {pages} {152701} (\bibinfo {year} {2021})}\BibitemShut {NoStop}%
\bibitem [{\citenamefont {Li}\ \emph {et~al.}(2021)\citenamefont {Li},
  \citenamefont {Qi}, \citenamefont {Lugaro}, \citenamefont {López},
  \citenamefont {Karakas}, \citenamefont {den Hartogh}, \citenamefont {Gao},\
  and\ \citenamefont {Tang}}]{Kuo_Ang_Li_2021_ApJL}%
  \BibitemOpen
  \bibfield  {author} {\bibinfo {author} {\bibfnamefont {K.-A.}\ \bibnamefont
  {Li}}, \bibinfo {author} {\bibfnamefont {C.}~\bibnamefont {Qi}}, \bibinfo
  {author} {\bibfnamefont {M.}~\bibnamefont {Lugaro}}, \bibinfo {author}
  {\bibfnamefont {A.~Y.}\ \bibnamefont {López}}, \bibinfo {author}
  {\bibfnamefont {A.~I.}\ \bibnamefont {Karakas}}, \bibinfo {author}
  {\bibfnamefont {J.}~\bibnamefont {den Hartogh}}, \bibinfo {author}
  {\bibfnamefont {B.-S.}\ \bibnamefont {Gao}},\ and\ \bibinfo {author}
  {\bibfnamefont {X.-D.}\ \bibnamefont {Tang}},\ }\bibfield  {title} {\bibinfo
  {title} {The stellar $\beta$-decay rate of 134cs and its impact on the barium
  nucleosynthesis in the s-process},\ }\href
  {https://doi.org/10.3847/2041-8213/ac260f} {\bibfield  {journal} {\bibinfo
  {journal} {The Astrophysical Journal Letters}\ }\textbf {\bibinfo {volume}
  {919}},\ \bibinfo {pages} {L19} (\bibinfo {year} {2021})}\BibitemShut
  {NoStop}%
\bibitem [{\citenamefont {Hara}\ and\ \citenamefont {Sun}(1995)}]{PSM_review}%
  \BibitemOpen
  \bibfield  {author} {\bibinfo {author} {\bibfnamefont {K.}~\bibnamefont
  {Hara}}\ and\ \bibinfo {author} {\bibfnamefont {Y.}~\bibnamefont {Sun}},\
  }\bibfield  {title} {\bibinfo {title} {Projected shell model and high-spin
  spectroscopy},\ }\href@noop {} {\bibfield  {journal} {\bibinfo  {journal}
  {Int. J. Mod. Phys. E}\ }\textbf {\bibinfo {volume} {4}},\ \bibinfo {pages}
  {637} (\bibinfo {year} {1995})}\BibitemShut {NoStop}%
\bibitem [{\citenamefont {Sun}\ and\ \citenamefont
  {Feng}(1996)}]{Sun_1996_Phys_Rep}%
  \BibitemOpen
  \bibfield  {author} {\bibinfo {author} {\bibfnamefont {Y.}~\bibnamefont
  {Sun}}\ and\ \bibinfo {author} {\bibfnamefont {D.~H.}\ \bibnamefont {Feng}},\
  }\bibfield  {title} {\bibinfo {title} {High spin spectroscopy with the
  projected shell model},\ }\href
  {https://doi.org/https://doi.org/10.1016/0370-1573(95)00049-6} {\bibfield
  {journal} {\bibinfo  {journal} {Phys. Rep.}\ }\textbf {\bibinfo {volume}
  {264}},\ \bibinfo {pages} {375} (\bibinfo {year} {1996})}\BibitemShut
  {NoStop}%
\bibitem [{\citenamefont {Mizusaki}\ \emph {et~al.}(2013)\citenamefont
  {Mizusaki}, \citenamefont {Oi}, \citenamefont {Chen},\ and\ \citenamefont
  {Sun}}]{Mizusaki_2013_PLB}%
  \BibitemOpen
  \bibfield  {author} {\bibinfo {author} {\bibfnamefont {T.}~\bibnamefont
  {Mizusaki}}, \bibinfo {author} {\bibfnamefont {M.}~\bibnamefont {Oi}},
  \bibinfo {author} {\bibfnamefont {F.-Q.}\ \bibnamefont {Chen}},\ and\
  \bibinfo {author} {\bibfnamefont {Y.}~\bibnamefont {Sun}},\ }\bibfield
  {title} {\bibinfo {title} {Grassmann integral and
  $\text{B}$alian–$\text{B}$rézin decomposition in
  $\text{H}$artree-$\text{F}$ock-$\text{B}$ogoliubov matrix elements},\ }\href
  {https://doi.org/https://doi.org/10.1016/j.physletb.2013.07.005} {\bibfield
  {journal} {\bibinfo  {journal} {Phys. Lett. B}\ }\textbf {\bibinfo {volume}
  {725}},\ \bibinfo {pages} {175} (\bibinfo {year} {2013})}\BibitemShut
  {NoStop}%
\bibitem [{\citenamefont {Wang}\ \emph {et~al.}(2014)\citenamefont {Wang},
  \citenamefont {Chen}, \citenamefont {Mizusaki}, \citenamefont {Oi},\ and\
  \citenamefont {Sun}}]{LJWang_2014_PRC_Rapid}%
  \BibitemOpen
  \bibfield  {author} {\bibinfo {author} {\bibfnamefont {L.-J.}\ \bibnamefont
  {Wang}}, \bibinfo {author} {\bibfnamefont {F.-Q.}\ \bibnamefont {Chen}},
  \bibinfo {author} {\bibfnamefont {T.}~\bibnamefont {Mizusaki}}, \bibinfo
  {author} {\bibfnamefont {M.}~\bibnamefont {Oi}},\ and\ \bibinfo {author}
  {\bibfnamefont {Y.}~\bibnamefont {Sun}},\ }\bibfield  {title} {\bibinfo
  {title} {Toward extremes of angular momentum: Application of the pfaffian
  algorithm in realistic calculations},\ }\href
  {https://doi.org/10.1103/PhysRevC.90.011303} {\bibfield  {journal} {\bibinfo
  {journal} {Phys. Rev. C}\ }\textbf {\bibinfo {volume} {90}},\ \bibinfo
  {pages} {011303(R)} (\bibinfo {year} {2014})}\BibitemShut {NoStop}%
\bibitem [{\citenamefont {Wang}\ \emph {et~al.}(2016)\citenamefont {Wang},
  \citenamefont {Sun}, \citenamefont {Mizusaki}, \citenamefont {Oi},\ and\
  \citenamefont {Ghorui}}]{LJWang_2016_PRC}%
  \BibitemOpen
  \bibfield  {author} {\bibinfo {author} {\bibfnamefont {L.-J.}\ \bibnamefont
  {Wang}}, \bibinfo {author} {\bibfnamefont {Y.}~\bibnamefont {Sun}}, \bibinfo
  {author} {\bibfnamefont {T.}~\bibnamefont {Mizusaki}}, \bibinfo {author}
  {\bibfnamefont {M.}~\bibnamefont {Oi}},\ and\ \bibinfo {author}
  {\bibfnamefont {S.~K.}\ \bibnamefont {Ghorui}},\ }\bibfield  {title}
  {\bibinfo {title} {Reduction of collectivity at very high spins in
  $^{134}\mathrm{Nd}$: Expanding the projected-shell-model basis up to
  10-quasiparticle states},\ }\href
  {https://doi.org/10.1103/PhysRevC.93.034322} {\bibfield  {journal} {\bibinfo
  {journal} {Phys. Rev. C}\ }\textbf {\bibinfo {volume} {93}},\ \bibinfo
  {pages} {034322} (\bibinfo {year} {2016})}\BibitemShut {NoStop}%
\bibitem [{\citenamefont {Chen}\ and\ \citenamefont
  {Wang}(2022)}]{ZRChen_2022_PRC}%
  \BibitemOpen
  \bibfield  {author} {\bibinfo {author} {\bibfnamefont {Z.-R.}\ \bibnamefont
  {Chen}}\ and\ \bibinfo {author} {\bibfnamefont {L.-J.}\ \bibnamefont
  {Wang}},\ }\bibfield  {title} {\bibinfo {title} {Pfaffian formulation for
  matrix elements of three-body operators in multiple quasiparticle
  configurations},\ }\href {https://doi.org/10.1103/PhysRevC.105.034342}
  {\bibfield  {journal} {\bibinfo  {journal} {Phys. Rev. C}\ }\textbf {\bibinfo
  {volume} {105}},\ \bibinfo {pages} {034342} (\bibinfo {year}
  {2022})}\BibitemShut {NoStop}%
\bibitem [{\citenamefont {Wang}\ \emph {et~al.}(2022)\citenamefont {Wang},
  \citenamefont {Gao}, \citenamefont {Wang},\ and\ \citenamefont
  {Sun}}]{BLWang_2022_PRC}%
  \BibitemOpen
  \bibfield  {author} {\bibinfo {author} {\bibfnamefont {B.-L.}\ \bibnamefont
  {Wang}}, \bibinfo {author} {\bibfnamefont {F.}~\bibnamefont {Gao}}, \bibinfo
  {author} {\bibfnamefont {L.-J.}\ \bibnamefont {Wang}},\ and\ \bibinfo
  {author} {\bibfnamefont {Y.}~\bibnamefont {Sun}},\ }\bibfield  {title}
  {\bibinfo {title} {Effective and efficient algorithm for the wigner rotation
  matrix at high angular momenta},\ }\href
  {https://doi.org/10.1103/PhysRevC.106.054320} {\bibfield  {journal} {\bibinfo
   {journal} {Phys. Rev. C}\ }\textbf {\bibinfo {volume} {106}},\ \bibinfo
  {pages} {054320} (\bibinfo {year} {2022})}\BibitemShut {NoStop}%
\bibitem [{\citenamefont {Wang}\ \emph {et~al.}(2019)\citenamefont {Wang},
  \citenamefont {Dong}, \citenamefont {Chen},\ and\ \citenamefont
  {Sun}}]{LJWang_2019_JPG}%
  \BibitemOpen
  \bibfield  {author} {\bibinfo {author} {\bibfnamefont {L.-J.}\ \bibnamefont
  {Wang}}, \bibinfo {author} {\bibfnamefont {J.}~\bibnamefont {Dong}}, \bibinfo
  {author} {\bibfnamefont {F.-Q.}\ \bibnamefont {Chen}},\ and\ \bibinfo
  {author} {\bibfnamefont {Y.}~\bibnamefont {Sun}},\ }\bibfield  {title}
  {\bibinfo {title} {Projected shell model analysis of structural evolution and
  chaoticity in fast-rotating nuclei},\ }\href
  {https://doi.org/10.1088/1361-6471/ab33be} {\bibfield  {journal} {\bibinfo
  {journal} {Journal of Physics G: Nuclear and Particle Physics}\ }\textbf
  {\bibinfo {volume} {46}},\ \bibinfo {pages} {105102} (\bibinfo {year}
  {2019})}\BibitemShut {NoStop}%
\bibitem [{\citenamefont {Wang}\ \emph {et~al.}(2020)\citenamefont {Wang},
  \citenamefont {Chen},\ and\ \citenamefont {Sun}}]{LJWang_PLB_2020_chaos}%
  \BibitemOpen
  \bibfield  {author} {\bibinfo {author} {\bibfnamefont {L.-J.}\ \bibnamefont
  {Wang}}, \bibinfo {author} {\bibfnamefont {F.-Q.}\ \bibnamefont {Chen}},\
  and\ \bibinfo {author} {\bibfnamefont {Y.}~\bibnamefont {Sun}},\ }\bibfield
  {title} {\bibinfo {title} {Basis-dependent measures and analysis
  uncertainties in nuclear chaoticity},\ }\href
  {https://doi.org/https://doi.org/10.1016/j.physletb.2020.135676} {\bibfield
  {journal} {\bibinfo  {journal} {Phys. Lett. B}\ }\textbf {\bibinfo {volume}
  {808}},\ \bibinfo {pages} {135676} (\bibinfo {year} {2020})}\BibitemShut
  {NoStop}%
\bibitem [{\citenamefont {Petrache}\ \emph {et~al.}(2023)\citenamefont
  {Petrache}, \citenamefont {Uusitalo}, \citenamefont {Briscoe}, \citenamefont
  {Sullivan}, \citenamefont {Joss}, \citenamefont {Tann}, \citenamefont
  {Aktas}, \citenamefont {Alayed}, \citenamefont {Al-Aqeel}, \citenamefont
  {Astier}, \citenamefont {Badran}, \citenamefont {Cederwall}, \citenamefont
  {Delafosse}, \citenamefont {Ertoprak}, \citenamefont {Favier}, \citenamefont
  {Forsberg}, \citenamefont {Gins}, \citenamefont {Grahn}, \citenamefont
  {Greenlees}, \citenamefont {He}, \citenamefont {Heery}, \citenamefont
  {Hilton}, \citenamefont {Kalantan}, \citenamefont {Li}, \citenamefont
  {Jodidar}, \citenamefont {Julin}, \citenamefont {Juutinen}, \citenamefont
  {Leino}, \citenamefont {Lewis}, \citenamefont {Li}, \citenamefont {Li},
  \citenamefont {Luoma}, \citenamefont {Lv}, \citenamefont {McCarter},
  \citenamefont {Nathaniel}, \citenamefont {Ojala}, \citenamefont {Page},
  \citenamefont {Pakarinen}, \citenamefont {Papadakis}, \citenamefont {Parr},
  \citenamefont {Partanen}, \citenamefont {Paul}, \citenamefont {Rahkila},
  \citenamefont {Ruotsalainen}, \citenamefont {Sandzelius}, \citenamefont
  {Sar\'en}, \citenamefont {Smallcombe}, \citenamefont {Sorri}, \citenamefont
  {Szwec}, \citenamefont {Wang}, \citenamefont {Wang}, \citenamefont {Waring},
  \citenamefont {Xu}, \citenamefont {Zhang}, \citenamefont {Zhang},
  \citenamefont {Zheng},\ and\ \citenamefont {Zimba}}]{Petrache_PRC_2023}%
  \BibitemOpen
  \bibfield  {author} {\bibinfo {author} {\bibfnamefont {C.~M.}\ \bibnamefont
  {Petrache}}, \bibinfo {author} {\bibfnamefont {J.}~\bibnamefont {Uusitalo}},
  \bibinfo {author} {\bibfnamefont {A.~D.}\ \bibnamefont {Briscoe}}, \bibinfo
  {author} {\bibfnamefont {C.~M.}\ \bibnamefont {Sullivan}}, \bibinfo {author}
  {\bibfnamefont {D.~T.}\ \bibnamefont {Joss}}, \bibinfo {author}
  {\bibfnamefont {H.}~\bibnamefont {Tann}}, \bibinfo {author} {\bibfnamefont
  {O.}~\bibnamefont {Aktas}}, \bibinfo {author} {\bibfnamefont
  {B.}~\bibnamefont {Alayed}}, \bibinfo {author} {\bibfnamefont {M.~A.~M.}\
  \bibnamefont {Al-Aqeel}}, \bibinfo {author} {\bibfnamefont {A.}~\bibnamefont
  {Astier}}, \bibinfo {author} {\bibfnamefont {H.}~\bibnamefont {Badran}},
  \bibinfo {author} {\bibfnamefont {B.}~\bibnamefont {Cederwall}}, \bibinfo
  {author} {\bibfnamefont {C.}~\bibnamefont {Delafosse}}, \bibinfo {author}
  {\bibfnamefont {A.}~\bibnamefont {Ertoprak}}, \bibinfo {author}
  {\bibfnamefont {Z.}~\bibnamefont {Favier}}, \bibinfo {author} {\bibfnamefont
  {U.}~\bibnamefont {Forsberg}}, \bibinfo {author} {\bibfnamefont
  {W.}~\bibnamefont {Gins}}, \bibinfo {author} {\bibfnamefont {T.}~\bibnamefont
  {Grahn}}, \bibinfo {author} {\bibfnamefont {P.~T.}\ \bibnamefont
  {Greenlees}}, \bibinfo {author} {\bibfnamefont {X.~T.}\ \bibnamefont {He}},
  \bibinfo {author} {\bibfnamefont {J.}~\bibnamefont {Heery}}, \bibinfo
  {author} {\bibfnamefont {J.}~\bibnamefont {Hilton}}, \bibinfo {author}
  {\bibfnamefont {S.}~\bibnamefont {Kalantan}}, \bibinfo {author}
  {\bibfnamefont {R.}~\bibnamefont {Li}}, \bibinfo {author} {\bibfnamefont
  {P.~M.}\ \bibnamefont {Jodidar}}, \bibinfo {author} {\bibfnamefont
  {R.}~\bibnamefont {Julin}}, \bibinfo {author} {\bibfnamefont
  {S.}~\bibnamefont {Juutinen}}, \bibinfo {author} {\bibfnamefont
  {M.}~\bibnamefont {Leino}}, \bibinfo {author} {\bibfnamefont {M.~C.}\
  \bibnamefont {Lewis}}, \bibinfo {author} {\bibfnamefont {J.~G.}\ \bibnamefont
  {Li}}, \bibinfo {author} {\bibfnamefont {Z.~P.}\ \bibnamefont {Li}}, \bibinfo
  {author} {\bibfnamefont {M.}~\bibnamefont {Luoma}}, \bibinfo {author}
  {\bibfnamefont {B.~F.}\ \bibnamefont {Lv}}, \bibinfo {author} {\bibfnamefont
  {A.}~\bibnamefont {McCarter}}, \bibinfo {author} {\bibfnamefont
  {S.}~\bibnamefont {Nathaniel}}, \bibinfo {author} {\bibfnamefont
  {J.}~\bibnamefont {Ojala}}, \bibinfo {author} {\bibfnamefont {R.~D.}\
  \bibnamefont {Page}}, \bibinfo {author} {\bibfnamefont {J.}~\bibnamefont
  {Pakarinen}}, \bibinfo {author} {\bibfnamefont {P.}~\bibnamefont
  {Papadakis}}, \bibinfo {author} {\bibfnamefont {E.}~\bibnamefont {Parr}},
  \bibinfo {author} {\bibfnamefont {J.}~\bibnamefont {Partanen}}, \bibinfo
  {author} {\bibfnamefont {E.~S.}\ \bibnamefont {Paul}}, \bibinfo {author}
  {\bibfnamefont {P.}~\bibnamefont {Rahkila}}, \bibinfo {author} {\bibfnamefont
  {P.}~\bibnamefont {Ruotsalainen}}, \bibinfo {author} {\bibfnamefont
  {M.}~\bibnamefont {Sandzelius}}, \bibinfo {author} {\bibfnamefont
  {J.}~\bibnamefont {Sar\'en}}, \bibinfo {author} {\bibfnamefont
  {J.}~\bibnamefont {Smallcombe}}, \bibinfo {author} {\bibfnamefont
  {J.}~\bibnamefont {Sorri}}, \bibinfo {author} {\bibfnamefont
  {S.}~\bibnamefont {Szwec}}, \bibinfo {author} {\bibfnamefont {L.~J.}\
  \bibnamefont {Wang}}, \bibinfo {author} {\bibfnamefont {Y.}~\bibnamefont
  {Wang}}, \bibinfo {author} {\bibfnamefont {L.}~\bibnamefont {Waring}},
  \bibinfo {author} {\bibfnamefont {F.~R.}\ \bibnamefont {Xu}}, \bibinfo
  {author} {\bibfnamefont {J.}~\bibnamefont {Zhang}}, \bibinfo {author}
  {\bibfnamefont {Z.~H.}\ \bibnamefont {Zhang}}, \bibinfo {author}
  {\bibfnamefont {K.~K.}\ \bibnamefont {Zheng}},\ and\ \bibinfo {author}
  {\bibfnamefont {G.}~\bibnamefont {Zimba}},\ }\bibfield  {title} {\bibinfo
  {title} {High-$k$ three-quasiparticle isomers in the proton-rich nucleus
  $^{129}\mathrm{Nd}$},\ }\href {https://doi.org/10.1103/PhysRevC.108.014317}
  {\bibfield  {journal} {\bibinfo  {journal} {Phys. Rev. C}\ }\textbf {\bibinfo
  {volume} {108}},\ \bibinfo {pages} {014317} (\bibinfo {year}
  {2023})}\BibitemShut {NoStop}%
\bibitem [{\citenamefont {Gao}\ \emph {et~al.}(2006)\citenamefont {Gao},
  \citenamefont {Sun},\ and\ \citenamefont {Chen}}]{Z_C_Gao_2006_GT}%
  \BibitemOpen
  \bibfield  {author} {\bibinfo {author} {\bibfnamefont {Z.-C.}\ \bibnamefont
  {Gao}}, \bibinfo {author} {\bibfnamefont {Y.}~\bibnamefont {Sun}},\ and\
  \bibinfo {author} {\bibfnamefont {Y.-S.}\ \bibnamefont {Chen}},\ }\bibfield
  {title} {\bibinfo {title} {Shell model method for gamow-teller transitions in
  heavy, deformed nuclei},\ }\href {https://doi.org/10.1103/PhysRevC.74.054303}
  {\bibfield  {journal} {\bibinfo  {journal} {Phys. Rev. C}\ }\textbf {\bibinfo
  {volume} {74}},\ \bibinfo {pages} {054303} (\bibinfo {year}
  {2006})}\BibitemShut {NoStop}%
\bibitem [{\citenamefont {Wang}\ \emph
  {et~al.}(2018{\natexlab{a}})\citenamefont {Wang}, \citenamefont {Sun},\ and\
  \citenamefont {Ghorui}}]{LJWang_2018_PRC_GT}%
  \BibitemOpen
  \bibfield  {author} {\bibinfo {author} {\bibfnamefont {L.-J.}\ \bibnamefont
  {Wang}}, \bibinfo {author} {\bibfnamefont {Y.}~\bibnamefont {Sun}},\ and\
  \bibinfo {author} {\bibfnamefont {S.~K.}\ \bibnamefont {Ghorui}},\ }\bibfield
   {title} {\bibinfo {title} {Shell-model method for gamow-teller transitions
  in heavy deformed odd-mass nuclei},\ }\href
  {https://doi.org/10.1103/PhysRevC.97.044302} {\bibfield  {journal} {\bibinfo
  {journal} {Phys. Rev. C}\ }\textbf {\bibinfo {volume} {97}},\ \bibinfo
  {pages} {044302} (\bibinfo {year} {2018}{\natexlab{a}})}\BibitemShut
  {NoStop}%
\bibitem [{\citenamefont {Tan}\ \emph {et~al.}(2020)\citenamefont {Tan},
  \citenamefont {Liu}, \citenamefont {Wang}, \citenamefont {Li},\ and\
  \citenamefont {Sun}}]{LJWang_PLB_2020_ec}%
  \BibitemOpen
  \bibfield  {author} {\bibinfo {author} {\bibfnamefont {L.}~\bibnamefont
  {Tan}}, \bibinfo {author} {\bibfnamefont {Y.-X.}\ \bibnamefont {Liu}},
  \bibinfo {author} {\bibfnamefont {L.-J.}\ \bibnamefont {Wang}}, \bibinfo
  {author} {\bibfnamefont {Z.}~\bibnamefont {Li}},\ and\ \bibinfo {author}
  {\bibfnamefont {Y.}~\bibnamefont {Sun}},\ }\bibfield  {title} {\bibinfo
  {title} {A novel method for stellar electron-capture rates of excited nuclear
  states},\ }\href
  {https://doi.org/https://doi.org/10.1016/j.physletb.2020.135432} {\bibfield
  {journal} {\bibinfo  {journal} {Phys. Lett. B}\ }\textbf {\bibinfo {volume}
  {805}},\ \bibinfo {pages} {135432} (\bibinfo {year} {2020})}\BibitemShut
  {NoStop}%
\bibitem [{\citenamefont {Wang}\ \emph
  {et~al.}(2021{\natexlab{a}})\citenamefont {Wang}, \citenamefont {Tan},
  \citenamefont {Li}, \citenamefont {Misch},\ and\ \citenamefont
  {Sun}}]{LJWang_2021_PRL}%
  \BibitemOpen
  \bibfield  {author} {\bibinfo {author} {\bibfnamefont {L.-J.}\ \bibnamefont
  {Wang}}, \bibinfo {author} {\bibfnamefont {L.}~\bibnamefont {Tan}}, \bibinfo
  {author} {\bibfnamefont {Z.}~\bibnamefont {Li}}, \bibinfo {author}
  {\bibfnamefont {G.~W.}\ \bibnamefont {Misch}},\ and\ \bibinfo {author}
  {\bibfnamefont {Y.}~\bibnamefont {Sun}},\ }\bibfield  {title} {\bibinfo
  {title} {Urca cooling in neutron star crusts and oceans: Effects of nuclear
  excitations},\ }\href {https://doi.org/10.1103/PhysRevLett.127.172702}
  {\bibfield  {journal} {\bibinfo  {journal} {Phys. Rev. Lett.}\ }\textbf
  {\bibinfo {volume} {127}},\ \bibinfo {pages} {172702} (\bibinfo {year}
  {2021}{\natexlab{a}})}\BibitemShut {NoStop}%
\bibitem [{\citenamefont {Wang}\ \emph
  {et~al.}(2021{\natexlab{b}})\citenamefont {Wang}, \citenamefont {Tan},
  \citenamefont {Li}, \citenamefont {Gao},\ and\ \citenamefont
  {Sun}}]{LJWang_2021_PRC_93Nb}%
  \BibitemOpen
  \bibfield  {author} {\bibinfo {author} {\bibfnamefont {L.-J.}\ \bibnamefont
  {Wang}}, \bibinfo {author} {\bibfnamefont {L.}~\bibnamefont {Tan}}, \bibinfo
  {author} {\bibfnamefont {Z.}~\bibnamefont {Li}}, \bibinfo {author}
  {\bibfnamefont {B.}~\bibnamefont {Gao}},\ and\ \bibinfo {author}
  {\bibfnamefont {Y.}~\bibnamefont {Sun}},\ }\bibfield  {title} {\bibinfo
  {title} {Description of $^{93}\mathrm{Nb}$ stellar electron-capture rates by
  the projected shell model},\ }\href
  {https://doi.org/10.1103/PhysRevC.104.064323} {\bibfield  {journal} {\bibinfo
   {journal} {Phys. Rev. C}\ }\textbf {\bibinfo {volume} {104}},\ \bibinfo
  {pages} {064323} (\bibinfo {year} {2021}{\natexlab{b}})}\BibitemShut
  {NoStop}%
\bibitem [{\citenamefont {Chen}\ and\ \citenamefont
  {Wang}(2023)}]{zrchen2023symm}%
  \BibitemOpen
  \bibfield  {author} {\bibinfo {author} {\bibfnamefont {Z.-R.}\ \bibnamefont
  {Chen}}\ and\ \bibinfo {author} {\bibfnamefont {L.-J.}\ \bibnamefont
  {Wang}},\ }\bibfield  {title} {\bibinfo {title} {Stellar $\beta$- decay rates
  for $^{63}\text{Co}$ and $^{63}\text{Ni}$ by the projected shell model},\
  }\href {https://doi.org/10.3390/sym15020315} {\bibfield  {journal} {\bibinfo
  {journal} {Symmetry}\ }\textbf {\bibinfo {volume} {15}},\ \bibinfo {pages}
  {315} (\bibinfo {year} {2023})}\BibitemShut {NoStop}%
\bibitem [{\citenamefont {Chen}\ and\ \citenamefont
  {Wang}(2024)}]{ZRChen_PLB2023}%
  \BibitemOpen
  \bibfield  {author} {\bibinfo {author} {\bibfnamefont {Z.-R.}\ \bibnamefont
  {Chen}}\ and\ \bibinfo {author} {\bibfnamefont {L.-J.}\ \bibnamefont
  {Wang}},\ }\bibfield  {title} {\bibinfo {title} {Stellar weak-interaction
  rates for $rp$-process waiting-point nuclei from projected shell model},\
  }\href {https://doi.org/https://doi.org/10.1016/j.physletb.2023.138338}
  {\bibfield  {journal} {\bibinfo  {journal} {Phys. Lett. B}\ }\textbf
  {\bibinfo {volume} {848}},\ \bibinfo {pages} {138338} (\bibinfo {year}
  {2024})}\BibitemShut {NoStop}%
\bibitem [{\citenamefont {Gao}\ \emph {et~al.}(2023)\citenamefont {Gao},
  \citenamefont {Chen},\ and\ \citenamefont {Wang}}]{FGao_PRC2023}%
  \BibitemOpen
  \bibfield  {author} {\bibinfo {author} {\bibfnamefont {F.}~\bibnamefont
  {Gao}}, \bibinfo {author} {\bibfnamefont {Z.-R.}\ \bibnamefont {Chen}},\ and\
  \bibinfo {author} {\bibfnamefont {L.-J.}\ \bibnamefont {Wang}},\ }\bibfield
  {title} {\bibinfo {title} {Nuclear $\ensuremath{\beta}$ spectrum from the
  projected shell model: Allowed one-to-one transition},\ }\href
  {https://doi.org/10.1103/PhysRevC.108.054313} {\bibfield  {journal} {\bibinfo
   {journal} {Phys. Rev. C}\ }\textbf {\bibinfo {volume} {108}},\ \bibinfo
  {pages} {054313} (\bibinfo {year} {2023})}\BibitemShut {NoStop}%
\bibitem [{\citenamefont {Wang}\ and\ \citenamefont
  {Wang}(2024)}]{BLWang_1stF_2024}%
  \BibitemOpen
  \bibfield  {author} {\bibinfo {author} {\bibfnamefont {B.-L.}\ \bibnamefont
  {Wang}}\ and\ \bibinfo {author} {\bibfnamefont {L.-J.}\ \bibnamefont
  {Wang}},\ }\bibfield  {title} {\bibinfo {title} {First-forbidden transition
  of nuclear $\beta$ decay by projected shell model},\ }\href
  {https://doi.org/https://doi.org/10.1016/j.physletb.2024.138515} {\bibfield
  {journal} {\bibinfo  {journal} {Phys. Lett. B}\ }\textbf {\bibinfo {volume}
  {850}},\ \bibinfo {pages} {138515} (\bibinfo {year} {2024})}\BibitemShut
  {NoStop}%
\bibitem [{\citenamefont {Hardy}\ and\ \citenamefont
  {Towner}(2009)}]{Hardy_2009_PRC}%
  \BibitemOpen
  \bibfield  {author} {\bibinfo {author} {\bibfnamefont {J.~C.}\ \bibnamefont
  {Hardy}}\ and\ \bibinfo {author} {\bibfnamefont {I.~S.}\ \bibnamefont
  {Towner}},\ }\bibfield  {title} {\bibinfo {title} {Superallowed
  ${0}^{+}\ensuremath{\rightarrow}{0}^{+}$ nuclear $\ensuremath{\beta}$ decays:
  A new survey with precision tests of the conserved vector current hypothesis
  and the standard model},\ }\href {https://doi.org/10.1103/PhysRevC.79.055502}
  {\bibfield  {journal} {\bibinfo  {journal} {Phys. Rev. C}\ }\textbf {\bibinfo
  {volume} {79}},\ \bibinfo {pages} {055502} (\bibinfo {year}
  {2009})}\BibitemShut {NoStop}%
\bibitem [{\citenamefont {Schenter}\ and\ \citenamefont
  {Vogel}(1983)}]{Fermi_func_1983}%
  \BibitemOpen
  \bibfield  {author} {\bibinfo {author} {\bibfnamefont {G.~K.}\ \bibnamefont
  {Schenter}}\ and\ \bibinfo {author} {\bibfnamefont {P.}~\bibnamefont
  {Vogel}},\ }\bibfield  {title} {\bibinfo {title} {A simple approximation of
  the fermi function in nuclear beta decay},\ }\href
  {https://doi.org/10.13182/NSE83-A17574} {\bibfield  {journal} {\bibinfo
  {journal} {Nuclear Science and Engineering}\ }\textbf {\bibinfo {volume}
  {83}},\ \bibinfo {pages} {393} (\bibinfo {year} {1983})},\ \Eprint
  {https://arxiv.org/abs/https://doi.org/10.13182/NSE83-A17574}
  {https://doi.org/10.13182/NSE83-A17574} \BibitemShut {NoStop}%
\bibitem [{\citenamefont {Zhi}\ \emph {et~al.}(2013)\citenamefont {Zhi},
  \citenamefont {Caurier}, \citenamefont {Cuenca-Garc\'{\i}a}, \citenamefont
  {Langanke}, \citenamefont {Mart\'{\i}nez-Pinedo},\ and\ \citenamefont
  {Sieja}}]{Zhi_FF_PRC_2013}%
  \BibitemOpen
  \bibfield  {author} {\bibinfo {author} {\bibfnamefont {Q.}~\bibnamefont
  {Zhi}}, \bibinfo {author} {\bibfnamefont {E.}~\bibnamefont {Caurier}},
  \bibinfo {author} {\bibfnamefont {J.~J.}\ \bibnamefont {Cuenca-Garc\'{\i}a}},
  \bibinfo {author} {\bibfnamefont {K.}~\bibnamefont {Langanke}}, \bibinfo
  {author} {\bibfnamefont {G.}~\bibnamefont {Mart\'{\i}nez-Pinedo}},\ and\
  \bibinfo {author} {\bibfnamefont {K.}~\bibnamefont {Sieja}},\ }\bibfield
  {title} {\bibinfo {title} {Shell-model half-lives including first-forbidden
  contributions for $r$-process waiting-point nuclei},\ }\href
  {https://doi.org/10.1103/PhysRevC.87.025803} {\bibfield  {journal} {\bibinfo
  {journal} {Phys. Rev. C}\ }\textbf {\bibinfo {volume} {87}},\ \bibinfo
  {pages} {025803} (\bibinfo {year} {2013})}\BibitemShut {NoStop}%
\bibitem [{\citenamefont {Cole}\ \emph {et~al.}(2012)\citenamefont {Cole},
  \citenamefont {Anderson}, \citenamefont {Zegers}, \citenamefont {Austin},
  \citenamefont {Brown}, \citenamefont {Valdez}, \citenamefont {Gupta},
  \citenamefont {Hitt},\ and\ \citenamefont {Fawwaz}}]{Cole_2012_PRC}%
  \BibitemOpen
  \bibfield  {author} {\bibinfo {author} {\bibfnamefont {A.~L.}\ \bibnamefont
  {Cole}}, \bibinfo {author} {\bibfnamefont {T.~S.}\ \bibnamefont {Anderson}},
  \bibinfo {author} {\bibfnamefont {R.~G.~T.}\ \bibnamefont {Zegers}}, \bibinfo
  {author} {\bibfnamefont {S.~M.}\ \bibnamefont {Austin}}, \bibinfo {author}
  {\bibfnamefont {B.~A.}\ \bibnamefont {Brown}}, \bibinfo {author}
  {\bibfnamefont {L.}~\bibnamefont {Valdez}}, \bibinfo {author} {\bibfnamefont
  {S.}~\bibnamefont {Gupta}}, \bibinfo {author} {\bibfnamefont {G.~W.}\
  \bibnamefont {Hitt}},\ and\ \bibinfo {author} {\bibfnamefont
  {O.}~\bibnamefont {Fawwaz}},\ }\bibfield  {title} {\bibinfo {title}
  {Gamow-teller strengths and electron-capture rates for $pf$-shell nuclei of
  relevance for late stellar evolution},\ }\href
  {https://doi.org/10.1103/PhysRevC.86.015809} {\bibfield  {journal} {\bibinfo
  {journal} {Phys. Rev. C}\ }\textbf {\bibinfo {volume} {86}},\ \bibinfo
  {pages} {015809} (\bibinfo {year} {2012})}\BibitemShut {NoStop}%
\bibitem [{\citenamefont {Sarriguren}(2013)}]{Sarriguren_2013_PRC}%
  \BibitemOpen
  \bibfield  {author} {\bibinfo {author} {\bibfnamefont {P.}~\bibnamefont
  {Sarriguren}},\ }\bibfield  {title} {\bibinfo {title} {Stellar
  electron-capture rates in $pf$-shell nuclei from quasiparticle random-phase
  approximation calculations},\ }\href
  {https://doi.org/10.1103/PhysRevC.87.045801} {\bibfield  {journal} {\bibinfo
  {journal} {Phys. Rev. C}\ }\textbf {\bibinfo {volume} {87}},\ \bibinfo
  {pages} {045801} (\bibinfo {year} {2013})}\BibitemShut {NoStop}%
\bibitem [{\citenamefont {Mart\'{\i}nez-Pinedo}\ \emph
  {et~al.}(2014)\citenamefont {Mart\'{\i}nez-Pinedo}, \citenamefont {Lam},
  \citenamefont {Langanke}, \citenamefont {Zegers},\ and\ \citenamefont
  {Sullivan}}]{Martinez_Pinedo_2014_PRC}%
  \BibitemOpen
  \bibfield  {author} {\bibinfo {author} {\bibfnamefont {G.}~\bibnamefont
  {Mart\'{\i}nez-Pinedo}}, \bibinfo {author} {\bibfnamefont {Y.~H.}\
  \bibnamefont {Lam}}, \bibinfo {author} {\bibfnamefont {K.}~\bibnamefont
  {Langanke}}, \bibinfo {author} {\bibfnamefont {R.~G.~T.}\ \bibnamefont
  {Zegers}},\ and\ \bibinfo {author} {\bibfnamefont {C.}~\bibnamefont
  {Sullivan}},\ }\bibfield  {title} {\bibinfo {title} {Astrophysical
  weak-interaction rates for selected $a=20$ and $a=24$ nuclei},\ }\href
  {https://doi.org/10.1103/PhysRevC.89.045806} {\bibfield  {journal} {\bibinfo
  {journal} {Phys. Rev. C}\ }\textbf {\bibinfo {volume} {89}},\ \bibinfo
  {pages} {045806} (\bibinfo {year} {2014})}\BibitemShut {NoStop}%
\bibitem [{\citenamefont {Brown}\ and\ \citenamefont
  {Wildenthal}(1985)}]{A.brown1985}%
  \BibitemOpen
  \bibfield  {author} {\bibinfo {author} {\bibfnamefont {B.~A.}\ \bibnamefont
  {Brown}}\ and\ \bibinfo {author} {\bibfnamefont {B.~H.}\ \bibnamefont
  {Wildenthal}},\ }\bibfield  {title} {\bibinfo {title} {Experimental and
  theoretical gamow-teller beta-decay observables for the sd-shell nuclei},\
  }\href@noop {} {\bibfield  {journal} {\bibinfo  {journal} {Atomic Data and
  Nuclear Data Tables}\ }\textbf {\bibinfo {volume} {33}},\ \bibinfo {pages}
  {347} (\bibinfo {year} {1985})}\BibitemShut {NoStop}%
\bibitem [{\citenamefont {Mart\'{\i}nez-Pinedo}\ \emph
  {et~al.}(1996)\citenamefont {Mart\'{\i}nez-Pinedo}, \citenamefont {Poves},
  \citenamefont {Caurier},\ and\ \citenamefont {Zuker}}]{martinez1996}%
  \BibitemOpen
  \bibfield  {author} {\bibinfo {author} {\bibfnamefont {G.}~\bibnamefont
  {Mart\'{\i}nez-Pinedo}}, \bibinfo {author} {\bibfnamefont {A.}~\bibnamefont
  {Poves}}, \bibinfo {author} {\bibfnamefont {E.}~\bibnamefont {Caurier}},\
  and\ \bibinfo {author} {\bibfnamefont {A.~P.}\ \bibnamefont {Zuker}},\
  }\bibfield  {title} {\bibinfo {title} {Effective ${g}_{A}$ in the
  $\mathrm{pf}$ shell},\ }\href {https://doi.org/10.1103/PhysRevC.53.R2602}
  {\bibfield  {journal} {\bibinfo  {journal} {Phys. Rev. C}\ }\textbf {\bibinfo
  {volume} {53}},\ \bibinfo {pages} {R2602} (\bibinfo {year}
  {1996})}\BibitemShut {NoStop}%
\bibitem [{\citenamefont {Gysbers}\ \emph {et~al.}(2019)\citenamefont
  {Gysbers}, \citenamefont {Hagen}, \citenamefont {Holt}, \citenamefont
  {Jansen}, \citenamefont {Morris}, \citenamefont {Navr\'atil}, \citenamefont
  {Papenbrock}, \citenamefont {Quaglioni}, \citenamefont {Schwenk},
  \citenamefont {Stroberg},\ and\ \citenamefont
  {Wendt}}]{Gysbers_2019_Nat_Phys}%
  \BibitemOpen
  \bibfield  {author} {\bibinfo {author} {\bibfnamefont {P.}~\bibnamefont
  {Gysbers}}, \bibinfo {author} {\bibfnamefont {G.}~\bibnamefont {Hagen}},
  \bibinfo {author} {\bibfnamefont {J.~D.}\ \bibnamefont {Holt}}, \bibinfo
  {author} {\bibfnamefont {G.~R.}\ \bibnamefont {Jansen}}, \bibinfo {author}
  {\bibfnamefont {T.~D.}\ \bibnamefont {Morris}}, \bibinfo {author}
  {\bibfnamefont {P.}~\bibnamefont {Navr\'atil}}, \bibinfo {author}
  {\bibfnamefont {T.}~\bibnamefont {Papenbrock}}, \bibinfo {author}
  {\bibfnamefont {S.}~\bibnamefont {Quaglioni}}, \bibinfo {author}
  {\bibfnamefont {A.}~\bibnamefont {Schwenk}}, \bibinfo {author} {\bibfnamefont
  {S.~R.}\ \bibnamefont {Stroberg}},\ and\ \bibinfo {author} {\bibfnamefont
  {K.~A.}\ \bibnamefont {Wendt}},\ }\bibfield  {title} {\bibinfo {title}
  {Discrepancy between experimental and theoretical $\beta$-decay rates
  resolved from first principles},\ }\href
  {https://doi.org/10.1038/s41567-019-0450-7} {\bibfield  {journal} {\bibinfo
  {journal} {Nature Physics}\ }\textbf {\bibinfo {volume} {15}},\ \bibinfo
  {pages} {428} (\bibinfo {year} {2019})}\BibitemShut {NoStop}%
\bibitem [{\citenamefont {Men\'endez}\ \emph {et~al.}(2011)\citenamefont
  {Men\'endez}, \citenamefont {Gazit},\ and\ \citenamefont
  {Schwenk}}]{Javier2011PRL}%
  \BibitemOpen
  \bibfield  {author} {\bibinfo {author} {\bibfnamefont {J.}~\bibnamefont
  {Men\'endez}}, \bibinfo {author} {\bibfnamefont {D.}~\bibnamefont {Gazit}},\
  and\ \bibinfo {author} {\bibfnamefont {A.}~\bibnamefont {Schwenk}},\
  }\bibfield  {title} {\bibinfo {title} {Chiral two-body currents in nuclei:
  Gamow-teller transitions and neutrinoless double-beta decay},\ }\href
  {https://doi.org/10.1103/PhysRevLett.107.062501} {\bibfield  {journal}
  {\bibinfo  {journal} {Phys. Rev. Lett.}\ }\textbf {\bibinfo {volume} {107}},\
  \bibinfo {pages} {062501} (\bibinfo {year} {2011})}\BibitemShut {NoStop}%
\bibitem [{\citenamefont {Wang}\ \emph
  {et~al.}(2018{\natexlab{b}})\citenamefont {Wang}, \citenamefont {Engel},\
  and\ \citenamefont {Yao}}]{LJWang_current_2018_Rapid}%
  \BibitemOpen
  \bibfield  {author} {\bibinfo {author} {\bibfnamefont {L.-J.}\ \bibnamefont
  {Wang}}, \bibinfo {author} {\bibfnamefont {J.}~\bibnamefont {Engel}},\ and\
  \bibinfo {author} {\bibfnamefont {J.~M.}\ \bibnamefont {Yao}},\ }\bibfield
  {title} {\bibinfo {title} {Quenching of nuclear matrix elements for
  $0\ensuremath{\nu}\ensuremath{\beta}\ensuremath{\beta}$ decay by chiral
  two-body currents},\ }\href {https://doi.org/10.1103/PhysRevC.98.031301}
  {\bibfield  {journal} {\bibinfo  {journal} {Phys. Rev. C}\ }\textbf {\bibinfo
  {volume} {98}},\ \bibinfo {pages} {031301(R)} (\bibinfo {year}
  {2018}{\natexlab{b}})}\BibitemShut {NoStop}%
\bibitem [{\citenamefont {M\"arkisch}\ \emph {et~al.}(2019)\citenamefont
  {M\"arkisch}, \citenamefont {Mest}, \citenamefont {Saul}, \citenamefont
  {Wang}, \citenamefont {Abele}, \citenamefont {Dubbers}, \citenamefont
  {Klopf}, \citenamefont {Petoukhov}, \citenamefont {Roick}, \citenamefont
  {Soldner},\ and\ \citenamefont {Werder}}]{arkisch2019}%
  \BibitemOpen
  \bibfield  {author} {\bibinfo {author} {\bibfnamefont {B.}~\bibnamefont
  {M\"arkisch}}, \bibinfo {author} {\bibfnamefont {H.}~\bibnamefont {Mest}},
  \bibinfo {author} {\bibfnamefont {H.}~\bibnamefont {Saul}}, \bibinfo {author}
  {\bibfnamefont {X.}~\bibnamefont {Wang}}, \bibinfo {author} {\bibfnamefont
  {H.}~\bibnamefont {Abele}}, \bibinfo {author} {\bibfnamefont
  {D.}~\bibnamefont {Dubbers}}, \bibinfo {author} {\bibfnamefont
  {M.}~\bibnamefont {Klopf}}, \bibinfo {author} {\bibfnamefont
  {A.}~\bibnamefont {Petoukhov}}, \bibinfo {author} {\bibfnamefont
  {C.}~\bibnamefont {Roick}}, \bibinfo {author} {\bibfnamefont
  {T.}~\bibnamefont {Soldner}},\ and\ \bibinfo {author} {\bibfnamefont
  {D.}~\bibnamefont {Werder}},\ }\bibfield  {title} {\bibinfo {title}
  {Measurement of the weak axial-vector coupling constant in the decay of free
  neutrons using a pulsed cold neutron beam},\ }\href
  {https://doi.org/10.1103/PhysRevLett.122.242501} {\bibfield  {journal}
  {\bibinfo  {journal} {Phys. Rev. Lett.}\ }\textbf {\bibinfo {volume} {122}},\
  \bibinfo {pages} {242501} (\bibinfo {year} {2019})}\BibitemShut {NoStop}%
\bibitem [{\citenamefont {Weidenm\"uller}(1961)}]{Weidenmuller_FF_RMP_1961}%
  \BibitemOpen
  \bibfield  {author} {\bibinfo {author} {\bibfnamefont {H.~A.}\ \bibnamefont
  {Weidenm\"uller}},\ }\bibfield  {title} {\bibinfo {title} {First-forbidden
  beta decay},\ }\href {https://doi.org/10.1103/RevModPhys.33.574} {\bibfield
  {journal} {\bibinfo  {journal} {Rev. Mod. Phys.}\ }\textbf {\bibinfo {volume}
  {33}},\ \bibinfo {pages} {574} (\bibinfo {year} {1961})}\BibitemShut
  {NoStop}%
\bibitem [{\citenamefont {Mougeot}(2015)}]{Mougeot_PRC_2015}%
  \BibitemOpen
  \bibfield  {author} {\bibinfo {author} {\bibfnamefont {X.}~\bibnamefont
  {Mougeot}},\ }\bibfield  {title} {\bibinfo {title} {Reliability of usual
  assumptions in the calculation of $\ensuremath{\beta}$ and $\ensuremath{\nu}$
  spectra},\ }\href {https://doi.org/10.1103/PhysRevC.91.055504} {\bibfield
  {journal} {\bibinfo  {journal} {Phys. Rev. C}\ }\textbf {\bibinfo {volume}
  {91}},\ \bibinfo {pages} {055504} (\bibinfo {year} {2015})}\BibitemShut
  {NoStop}%
\bibitem [{\citenamefont {Haaranen}\ \emph {et~al.}(2017)\citenamefont
  {Haaranen}, \citenamefont {Kotila},\ and\ \citenamefont
  {Suhonen}}]{Suhonen_PRC_2017_general_Fermi_for_unique}%
  \BibitemOpen
  \bibfield  {author} {\bibinfo {author} {\bibfnamefont {M.}~\bibnamefont
  {Haaranen}}, \bibinfo {author} {\bibfnamefont {J.}~\bibnamefont {Kotila}},\
  and\ \bibinfo {author} {\bibfnamefont {J.}~\bibnamefont {Suhonen}},\
  }\bibfield  {title} {\bibinfo {title} {Spectrum-shape method and the
  next-to-leading-order terms of the $\ensuremath{\beta}$-decay shape factor},\
  }\href {https://doi.org/10.1103/PhysRevC.95.024327} {\bibfield  {journal}
  {\bibinfo  {journal} {Phys. Rev. C}\ }\textbf {\bibinfo {volume} {95}},\
  \bibinfo {pages} {024327} (\bibinfo {year} {2017})}\BibitemShut {NoStop}%
\bibitem [{NND()}]{NNDC}%
  \BibitemOpen
  \href@noop {} {\bibinfo {title} {https://www.nndc.bnl.gov}}\BibitemShut
  {NoStop}%
\end{thebibliography}

%

\end{document}